 \documentclass[manuscript]{emulateapj-rtx4}

\usepackage{color}

\newcommand{\rhoGJ}{\rho_{{\rm GJ}}}  
\newcommand{\Ell}{E_\parallel}      
\newcommand{\rhowSQR}{\rho_{\rm w}^2}

\slugcomment{}

\shorttitle{Lepton acceleration in the vicinity of the event horizon}
\shortauthors{Hirotani et al.}

\begin{document}


\title{Lepton acceleration in the vicinity of the event horizon:
       Very-high-energy emissions from 
       super-massive black holes}


\author{}
  \author{Kouichi Hirotani${}^1$, 
          Hung-Yi Pu${}^1$, 
          Lupin Chun-Che Lin${}^1$,
          Albert K. H Kong${}^2$,
          Satoki Matsushita${}^1$, 
          Keiichi Asada${}^1$,
          Hsiang-Kuang Chang${}^2$, and
          Pak-Hin T. Tam${}^3$ 
          }
\affil{${}^1$
       Academia Sinica, Institute of Astronomy and Astrophysics (ASIAA),
       PO Box 23-141, Taipei, Taiwan 10617, R.O.C.;
       hirotani@tiara.sinica.edu.tw}
\affil{${}^2$
       Institute of Astronomy,
       Department of Physics, 
       National Tsing Hua University,
       No. 101, Section 2, Kuang-Fu Road, Hsinchu, Taiwan 30013, R.O.C.}
\affil{${}^3$
       School of Physics and Astronomy, Sun Yat-Sen University, 
       Zhuhai 519082, China}

%
%
%
%


\begin{abstract}
Around a rapidly rotating black hole (BH),
when the plasma accretion rate is much less than the Eddington rate,
the radiatively inefficient accretion flow (RIAF) 
cannot supply enough MeV photons that are capable of
materializing as pairs.
In such a charge-starved BH magnetosphere,
the force-free condition breaks down in the polar funnels.
Applying the pulsar outer-magnetospheric lepton accelerator theory
to super-massive BHs,
we demonstrate that 
a strong electric field arises along the magnetic field lines
in the direct vicinity of the event horizon in the funnels,
that the electrons and positrons are accelerated up to 100~TeV
in this vacuum gap,
and that these leptons emit
copious photons via inverse-Compton (IC) process
between 0.1~TeV and 30~TeV for a distant observer.
It is found that these IC fluxes will be detectable
with Imaging Atmospheric Cherenkov Telescopes,
provided that a low-luminosity active galactic nucleus is 
located within 1~Mpc for a million-solar-mass central BH
or within 30~Mpc for a billion-solar-mass central BH. 
These very-high-energy fluxes
are beamed in a relatively small solid angle around the rotation axis
because of the inhomogeneous and anisotropic distribution of 
the RIAF photon field,
and show an anti-correlation with the RIAF submillimeter fluxes.
The gap luminosity little depends on the three-dimensional
magnetic-field configuration, 
because the Goldreich-Julian charge density,
and hence the exerted electric field
is essentially governed by the frame-dragging effect,
not by the magnetic field configuration.
\end{abstract}





\section{Introduction}
\label{sec:intro}
%
%

It is commonly accepted that every active galaxy harbors a
super-massive black hole (SMBH) in its center
with a mass ranging typically in $10^6-10^{10} M_\odot$
\citep{miyoshi95,ferrarese96,remco16,larkin16}.
A likely mechanism for powering such an active galactic nucleus (AGN),
is the release of the gravitational energy of accreting plasmas
\citep{lynden69}
or the electromagnetic extraction 
of the rotational energy of a rotating SMBH.
The latter mechanism,
which is called the Blandford-Znajek (BZ) mechanism
\citep{bla76},
works only when there is a plasma accretion,
because the central black hole (BH) 
cannot have its own magnetic moment
\citep[e.g.,][]{MTW73}.
As long as the magnetic field energy is in a rough equipartition
with the gravitational binding energy of the accreting plasmas,
both mechanisms contribute comparably 
in terms of luminosity.
The former mechanism is supposed to power
the mildly relativistic winds that are launched from
the accretion disks 
\citep{meier01,hujeirat04,sadow10}.
There is, however, growing evidence that
relativistic jets are energized by the latter, BZ mechanism
through numerical simulations
\citep{koide02,mckinney06,mckinney12}
(see also \citealt[][]{punsly11} for an ergospheric disc jet model).
Indeed, general relativistic (GR) magnetohydrodynamic (MHD) models 
show the existence of collimated and magnetically dominated jets
in the polar regions
\citep{hirose04,mckinney04,tchek10},
whose structures are similar to those 
in the force-free models
\citep{hawley06,mckinney07a,mckinney07b}.
Since the centrifugal-force barrier prevents plasma accretion 
towards the rotation axis,
the magnetic energy density dominates the plasmas' 
rest-mass energy density in these polar funnels. 

Within such a nearly vacuum, polar funnel,
electron-positron pairs are supplied via 
the collisions of MeV photons emitted from 
the equatorial, accreting region.
For example, 
when the mass accretion rate is typically less than
1~\% of the Eddington rate, the accreting plasmas form 
a radiatively inefficient accretion flow (RIAF),
emitting radio to infrared photons via synchrotron process
and MeV photons via free-free and IC processes
\citep{ichimaru77,narayan94,narayan95,abram95,mahad97,
  esin97,esin98,bla99,manmoto00}.
Particularly, 
when the accretion rate becomes much less than the Eddington rate
\citep{levi11},
the RIAF MeV photons can no longer sustain 
a force-free magnetosphere,
which inevitably leads to an appearance of an electric field,
$E_\parallel$,
along the magnetic field lines in the polar funnel.
In such a vacuum gap, 
we can expect that the BZ power 
may be partially dissipated as particle
acceleration and emission,
in the same manner as in pulsar outer-gap models
\citep{hollo73,cheng86a,cheng86b,chiang92,romani96,cheng00,
 romani10,hiro13,tak04,tak16}. 

In this context, \citet{bes92} demonstrated that 
the Goldreich-Julian charge density vanishes
in the vicinity of the even horizon due to space-time frame dragging,
and that a vacuum gap does arise around this null-charge surface.
Subsequently, \citet{hiro98,nero07,levi11,brod15,hiro16}
extended this BH gap model to quantify its electrodynamics.
Within a BH gap, 
electrons and positrons,
which are referred to as leptons in the present paper, 
are created and accelerated 
into the opposite directions by $E_\parallel$
to emit copious $\gamma$-rays in high energies 
(HE, typically between 0.1~GeV and 100~GeV)
via the curvature process for stellar-mass BHs
and in very high energies 
(VHE, typically between 0.1~TeV and 100~TeV)
via the IC process for SMBHs.
Recently, \citet[][hereafter H16]{hiro16b}
examined the BH gap for various BH masses
and demonstrated that these HE and VHE fluxes are detectable
at Earth, provided that the BH is located close enough
and that the accretion rate is in a certain,
relatively narrow range.

In the present paper, 
to further quantify the BH gap model,
we consider an inhomogeneous and anisotropic RIAF photon field
in the polar funnel and explicitly solve the distribution functions
of the gap-accelerated leptons.
After describing the background space time
in \S~\ref{sec:geometry},
we focus on the RIAF photon field in \S~\ref{sec:soft}.
Then in \S~\ref{sec:accelerator},
we formulate the Poisson equation that describes $E_\parallel$,
the lepton Boltzmann equations, and
the radiative transfer equation of the emitted photons.
We show the results in \S~\ref{sec:solutions},
focusing on the particle distribution functions
and the resultant $\gamma$-ray spectra for SMBHs.
We finally compare the BH gaps with the pulsar outer gaps
in \S~\ref{sec:disc}.

\section{Background geometry}
\label{sec:geometry}
Let us start with
describing the background spacetime geometry.
We adopt the geometrized unit, putting $c=G=1$, where
$c$ and $G$ denote the speed of light and the gravitational constant,
respectively.
Around a rotating BH, 
the background geometry is described by the Kerr metric
\citep{kerr63}.
In the Boyer-Lindquist coordinates, it becomes 
\citep{boyer67} 
\begin{equation}
 ds^2= g_{tt} dt^2
      +2g_{t\varphi} dt d\varphi
      +g_{\varphi\varphi} d\varphi^2
      +g_{rr} dr^2
      +g_{\theta\theta} d\theta^2,
  \label{eq:metric}
\end{equation}
where 
\begin{equation}
   g_{tt} 
   \equiv 
   -\frac{\Delta-a^2\sin^2\theta}{\Sigma},
   \qquad
   g_{t\varphi}
   \equiv 
   -\frac{2Mar \sin^2\theta}{\Sigma}, 
  \label{eq:metric_2}
\end{equation}
\begin{equation}
   g_{\varphi\varphi}
     \equiv 
     \frac{A \sin^2\theta}{\Sigma} , 
     \qquad
   g_{rr}
     \equiv 
     \frac{\Sigma}{\Delta} , 
     \qquad
   g_{\theta\theta}
     \equiv 
     \Sigma ;
  \label{eq:metric_3}
\end{equation}
$\Delta \equiv r^2-2Mr+a^2$,
$\Sigma\equiv r^2 +a^2\cos^2\theta$,
$A \equiv (r^2+a^2)^2-\Delta a^2\sin^2\theta$.
At the horizon, we obtain $\Delta=0$, 
which gives the horizon radius, 
$r_{\rm H} \equiv M+\sqrt{M^2-a^2}$,
where $M$ corresponds to the gravitational radius,
$r_{\rm g} \equiv GM c^{-2}$.
The spin parameter $a$ becomes $a=M$ for a maximally rotating BH,
and becomes $a=0$ for a non-rotating BH. 

We assume that the non-corotational potential $\Phi$
depends on $t$ and $\varphi$ only through
the form $\varphi-\Omega_{\rm F} t$, and put
\begin{equation}
  F_{\mu t}+\Omega_{\rm F} F_{\mu \varphi}
  = -\partial_\mu \Phi(r,\theta,\varphi-\Omega_{\rm F} t) ,
  \label{eq:def_Phi}
\end{equation}
where $\Omega_{\rm F}$ denotes the magnetic-field-line
rotational angular frequency.
We refer to such a solution as a \lq stationary' solution
in the present paper,
because the solution is unchanged in the corotating frame
of the magnetosphere.
Note that the solution is valid not only between the
two light surfaces (i.e., where 
$k_0 \equiv -g_{tt}-2g_{t\varphi}\Omega_{\rm F}
           -g_{\varphi\varphi}\Omega_{\rm F}^2 > 0$),
but also inside the inner light surface or outside the outer
light surface 
(i.e., where $k_0<0$ and the corotating motion becomes space-like)
\citep{znaj77,taka90}.
For example, 
the exact analytic solution of the electromagnetic field
in a striped pulsar wind is 
of this functional form
and is valid outside the light cylinder
\citep{bogov99,petri05,petri13}.
In another word, the toroidal velocity
of the magnetic field lines,
$\sqrt{g_{\varphi\varphi}} \, \Omega_{\rm F}$, 
is merely a phase velocity,
where $\sqrt{g_{\varphi\varphi}}$ denotes the distance
from the rotation axis.

The Gauss's law gives the Poisson equation
that describes $\Phi$ in a three dimensional magnetosphere
(eq.~[15] of H16),
\begin{equation}
  -\frac{1}{\sqrt{-g}}
   \partial_\mu 
      \left( \frac{\sqrt{-g}}{\rhowSQR}
             g^{\mu\nu} g_{\varphi\varphi}
             \partial_\nu \Phi
      \right)
  = 4\pi(\rho-\rhoGJ),
  \label{eq:pois}
\end{equation}
where the GR Goldreich-Julian (GJ) charge density
is defined as
\citep{GJ69,mestel71,hiro06b}
\begin{equation}
  \rhoGJ \equiv 
      \frac{1}{4\pi\sqrt{-g}}
      \partial_\mu \left[ \frac{\sqrt{-g}}{\rhowSQR}
                         g^{\mu\nu} g_{\varphi\varphi}
                         (\Omega_{\rm F}-\omega) F_{\varphi\nu}
                 \right].
  \label{eq:def_GJ}
\end{equation}
If the real charge density $\rho$ deviates from the
rotationally induced Goldreich-Julian charge density,
$\rho_{\rm GJ}$, in some region,
equation~(\ref{eq:pois}) shows that
$\Phi$ changes as a function of position.
Thus, an acceleration electric field, 
$E_\parallel= -\partial \Phi / \partial s$,
arises along the magnetic field line,
where $s$ denotes the distance along the magnetic field line.
A gap is defined as the spatial region in which $E_\parallel$
is non-vanishing.
At the null charge surface,
$\rhoGJ$ changes sign by definition.
Thus, a vacuum gap, 
in which $\vert\rho\vert \ll \vert\rhoGJ\vert$,
appears around the null-charge surface,
because $\partial E_\parallel / \partial s$
should have opposite signs at the inner and outer boundaries
\citep{hollo73,chiang92,romani96,cheng00}. 
As an extension of the vacuum gap, 
a non-vacuum gap,
in which $\vert\rho\vert$ becomes a good fraction of 
$\vert\rhoGJ\vert$,
also appears around the null-charge surface
(\S~2.3.2 of HP~16),
unless the injected current across either the inner or the outer
boundary becomes a substantial fraction of the GJ value.

It should be noted that
$\rhoGJ$ vanishes (and hence the null surface appears)
near the place where $\Omega_{\rm F}$ coincides with 
the space-time dragging angular frequency, $\omega$ \citep{bes92}.
The deviation of the null surface
from this $\omega(r,\theta)=\Omega_{\rm F}$ surface is,
indeed, small, as figure~1 of \citet{hiro98} indicates.
Since $\omega$ matches $\Omega_{\rm F}$ only near the horizon,
the null surface, and hence the gap generally appears 
within one or two gravitational radii above the horizon,
irrespective of the BH mass. 
It is also noteworthy that $\Omega_{\rm F}$ little changes along 
the individual magnetic field lines,
because the force-free approximation breaks down only slightly
as will be shown in \S~\ref{sec:billion}
by a comparison between the potential drop and the electro-motive force.

\section{Propagation of soft photons}
\label{sec:soft}
To quantify the gap electrodynamics, 
we need to compute the pair creation rate. 
To this end, we must tabulate the specific intensity
of the soft photons at each position in the polar funnel.
In this section, we therefore
consider how the soft photons are emitted in a RIAF
and propagate around a rotating BH.
We assume that the soft photon field is axisymmetric
with respect to the BH rotation axis.

\subsection{Emission from equatorial region}
\label{sec:riaf}
When the mass accretion rate is much small compared to the
Eddington rate, the accreting plasmas form a RIAF 
with a certain thickness in the equatorial region. 
For simplicity, in this paper,
we approximate that the such plasmas
rotate around the BH with the general-relativistic Keplerian
angular velocity, 
\begin{equation}
  \Omega_{\rm K}
  \equiv \pm \frac{1}{M}
             \frac{1}{(r/M)^{3/2} \pm a/M},
  \label{eq:GR_Kepler}
\end{equation}
and that their motion is dominated by this rotation.
That is, we neglect the motion of 
the soft-photon-emitting plasmas on the poloidal plane,
($r$,$\theta$),
for simplicity.

Let us introduce the local rest frame (LRF) of such rotating plasmas.
Putting $\beta^\varphi=\Omega_{\rm K}$  
in equations~(\ref{eq:rotating_e0})--(\ref{eq:rotating_e2}),
we obtain its orthonormal tetrad,
\begin{equation}
  \mbox{\boldmath$e$}_{(\hat{t})}{}^{\tiny\rm LRF}
  = \left(\frac{dt}{d\tau}\right)^{\tiny\rm LRF}
    \left( \partial_t + \Omega_{\rm K} \partial_\varphi 
    \right), 
  \label{eq:LRF_e0o}
\end{equation}
\begin{equation}
  \mbox{\boldmath$e$}_{(\hat{\varphi})}{}^{\tiny\rm LRF}
  = \frac{g_{t\varphi}+g_{\varphi\varphi}\Omega_{\rm K}}
         {\rho_{\rm w} \sqrt{D}}
    \partial_t
   -\frac{g_{tt}+g_{t\varphi}\Omega_{\rm K}}
         {\rho_{\rm w} \sqrt{D}}
    \partial_\varphi, 
  \label{eq:LRF_e3o}
\end{equation}
\begin{equation}
  \mbox{\boldmath$e$}_{(\hat{r})}{}^{\tiny\rm LRF}
  =\sqrt{g^{rr}} \partial_r
  \label{eq:LRF_e1o}
\end{equation}
\begin{equation}
  \mbox{\boldmath$e$}_{(\hat{\theta})}{}^{\tiny\rm LRF}
  =\sqrt{g^{\theta\theta}} \partial_\theta ,
  \label{eq:LRF_e2o}
\end{equation}
where 
\begin{equation}
  D \equiv -g_{tt} -2g_{t\varphi} \Omega_{\rm K}
           -g_{\varphi\varphi} \Omega_{\rm K}{}^2,
  \label{eq:LRF_D}
\end{equation}
and the redshift factor becomes
\begin{equation}
  \left(\frac{d\tau}{dt}\right)^{\tiny\rm LRF}
   =\sqrt{D}.
  \label{eq:LRF_redshift}
\end{equation}

In LRF, the photon propagation direction is expressed
in terms of the photon wave vector, $k_\alpha$.
Denoting the photon energy as 
$-k_{\hat{t}}{}^{\tiny\rm LRF}=\omega_{\tiny\rm LRF}$ in LRF,
we obtain from the dispersion relation $k^\mu k_\mu=0$,
\begin{equation}
  (k_{\hat r}{}^{\tiny\rm LRF})^2
 +(k_{\hat \theta}{}^{\tiny\rm LRF})^2
 +(k_{\hat \varphi}{}^{\tiny\rm LRF})^2
 = \omega_{\tiny\rm LRF}^2,
  \label{eq:n_i}
\end{equation}
where
$k_{\hat\mu}{}^{\tiny\rm LRF}
 = k_\alpha \left[ \mbox{\boldmath$e$}_{(\hat\mu)}{}^{\tiny\rm LRF} 
         \right]^\alpha
$.
We therefore parameterize the direction of the photons 
emitted by the RIAF plasma
with two angles $\theta_\gamma{}^{\rm LRF}$ and 
$\varphi_\gamma{}^{\rm LRF}$ in LRF,
and put
\begin{eqnarray}
  k_{\hat r}{}^{\tiny\rm LRF}
  &=& \omega_{\tiny\rm LRF} \cos\theta_\gamma{}^{\tiny\rm LRF},
      \label{eq:LRF_n1}\\
  k_{\hat \theta}{}^{\tiny\rm LRF}
  &=& \omega_{\tiny\rm LRF} \sin\theta_\gamma{}^{\tiny\rm LRF}
      \cos\varphi_\gamma{}^{\tiny\rm LRF},
  \label{eq:LRF_n2}\\
  k_{\hat \varphi}{}^{\tiny\rm LRF}
  &=& \omega_{\tiny\rm LRF} \sin\theta_\gamma{}^{\tiny\rm LRF}
      \sin\varphi_\gamma{}^{\tiny\rm LRF}.
  \label{eq:LRF_n3}
\end{eqnarray}



\begin{figure}
 \includegraphics[angle=0,scale=1.0]{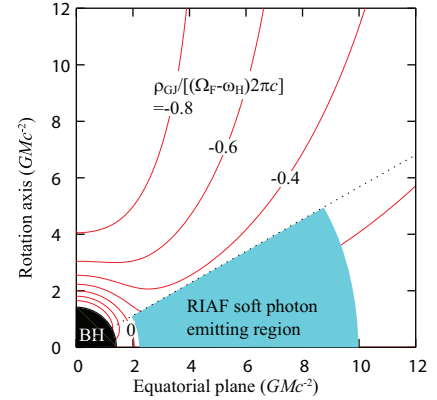}
\caption{
Side view of an axisymmetric black hole magnetosphere.
As a toy model, 
we assume that the radiatively inefficient accretion flow (RIAF) 
and the polar funnel is bounded by a cone (dotted line)
with colatitude $\theta=60^\circ$.
The RIAF photons are assumed to be emitted from the cyan
shaded region,
which is within the colatitude
$60^\circ < \theta < 90^\circ$
and the radius 
$r_{\tiny\rm ISCO}=2.31M < r < r_{\rm out}=10M$
in the Boyer-Lindquist coordinate.
The red curves denote the dimensionless Goldreich-Julian
(GJ) charge density,
where the magnetic field lines are assumed to be radial
on the poloidal plane 
(see also the left panel of fig.~1 in HP16).
The red curve labeled with \lq\lq 0''
(near $r=2M$) denotes the null-charge surface,
around which a gap arises 
in the polar funnel, $0 \le \theta<60^\circ$.
Due to frame dragging,
the GJ charge density increases inwards
near the horizon.
\label{fig:sideview}
}
\end{figure}

We assume that the soft photons are emitted 
homogeneously and isotropically in this LRF. 
In the configuration space,
we divide the soft-photon emission region into 
$14$ meridional bins 
between in $60^\circ < \theta < 90^\circ$ from the rotation axis,
and $14$ radial bins between 
$r_{\tiny\rm ISCO} < r < r_{\rm out}$, %
where $r_{\tiny\rm ISCO}$ denotes the radial Boyer-Lindquist coordinate
at the inner-most stable circular orbit (ISCO),
and $r_{\rm out}=10 M$.
We illustrate this RIAF-emitting region 
in figure~\ref{fig:sideview}.
For $a=0.9M$, we obtain 
$r_{\tiny\rm ISCO}=2.31M$ and
$r_{\rm H}=1.43M$.
We adopt the emission points
with a constant meridional interval 
$d\theta=(90^\circ-60^\circ)/14$.
To achieve a spatially homogeneous emission, 
we adopt a radial interval $dr$
so that the integration of the area
\begin{equation}
  S \equiv \int_r^{r+dr} 
           \int_\theta^{\theta+d\theta} 
             \sqrt{g_{rr} g_{\theta\theta}} dr d\theta
  \label{eq:LRF_S}
\end{equation}
may be constant.

In the momentum space,
we divide the emission direction of the photons
into 200 azimuthal-propagation-direction bins 
in $0 < \varphi_\gamma{}^{\tiny\rm LRF} < 2\pi$, and
128 latitudinal-propagation-direction bins
in $-1 < \cos\theta_\gamma{}^{\tiny\rm LRF} < 1$.
To achieve an isotropic emission,
we emit test photons isotropically
with a constant $\varphi_\gamma{}^{\tiny\rm LRF}$ interval
and a constant $\cos\theta_\gamma{}^{\tiny\rm LRF}$ interval.

Between the distant static observer (i.e., us)
and the LRF,
the photon energy changes by the redshift factor,
\begin{eqnarray}
  \frac{\omega_\infty}{\omega_{\tiny\rm LRF}}
  &=&
    \frac{\mbox{\boldmath$e$}_{(t)}{}^\infty \cdot
          \mbox{\boldmath$k$}}
         {\mbox{\boldmath$e$}_{(t)}{}^{\tiny\rm LRF} \cdot 
          \mbox{\boldmath$k$}}
  = \left(\frac{d\tau}{dt}\right)^{\tiny\rm LRF}
    \frac{-\hbar \omega_\infty}
         {(\mbox{\boldmath$e$}_{(t)}
           +\beta^\varphi\mbox{\boldmath$e$}_{(\varphi)})
          \cdot \mbox{\boldmath$k$}
         }
  \nonumber\\
  &=&
    \left(\frac{d\tau}{dt}\right)^{\tiny\rm LRF}
    \frac{1}{1-\beta^\varphi\lambda},
  \label{eq:LRF_redshift_ph}
\end{eqnarray}
where $\lambda \equiv k_\varphi / \hbar \omega_\infty$ 
denotes the ratio between the photon
angular momentum $k_\varphi$ and energy $-k_t=\hbar \omega_\infty$,
both of which are conserved along the ray.

In the coordinate basis, 
the photon wave vector takes the following components:
\begin{eqnarray}
  k_t 
  = -\hbar \omega_\infty
 &\equiv&
   k_\alpha \left[ \mbox{\boldmath$e$}_{(t)} \right]^\alpha
    = \left(\frac{d\tau}{dt}\right)^{\tiny\rm LRF}
      \frac{-\omega_{\tiny\rm LRF}}
           {1-\beta^\varphi\lambda} ,
    \label{eq:LRF_k0}\\
  k_\varphi 
  = \lambda \hbar \omega_\infty
  &\equiv&
    k_\alpha \left[ \mbox{\boldmath$e$}_{(\varphi)} \right]^\alpha ,
    \label{eq:LRF_k3}\\
  k_r
  &\equiv&
    k_\alpha \left[ \mbox{\boldmath$e$}_{(r)} \right]^\alpha
    = \sqrt{g_{rr}} k_{\hat{r}}{}^{\tiny\rm LRF} ,
   \label{eq:LRF_k1}\\
  k_\theta
  &\equiv&
    k_\alpha \left[ \mbox{\boldmath$e$}_{(\theta)} \right]^\alpha
    = \sqrt{g_{\theta\theta}} k_{\hat{\theta}}{}^{\tiny\rm LRF} ,
    \label{eq:LRF_k2}
\end{eqnarray}
where $\mbox{\boldmath$e$}_{(\mu)} \equiv \partial_\mu$.
In the next subsection.
we use these components, $k_\mu$,
to ray-trace the photons emitted in the LRF isotropically.

\subsection{Light propagation around the BH}
\label{sec:raytrace}
As a photon propagates, 
its energy $-k_t$, angular momentum $k_\varphi$, and
the Carter's constant \citep{carter68}
\begin{equation}
  Q 
  \equiv 
  (k_\theta/k_t)^2 + a^2 \cos^2\theta + \lambda^2\cot^2\theta
  \label{eq:Carter}
\end{equation}
are conserved along the ray.
Since $Q$ is finite,
only the photons having vanishing angular momenta $\lambda$
can propagate toward the rotation axis, $\theta \sim 0$.
It is worth noting that most of the soft photons have
positive angular momenta, because they are emitted by
the rotating plasmas in the RIAF.
Figure~\ref{fig:sideview} shows such a situation that soft photons
 are most efficiently emitted outside the ISCO, thereby having
 positive angular momenta except when they are emitted into 
 a specific counter-rotational direction in the LRF.
 As the disc angular momentum increases,
 the solid angle into which the photons
 with a fixed range of very small angular momenta 
 (in an absolute value sense) propagate, decreases in the LRF.
As a result, only a small portion of the soft photons can propagate
to $\theta \sim 0$,
leading to a smaller soft photon density 
in the higher latitudes, $\theta \ll 1$, 
compared to the middle latitudes, $\theta \sim 1$.
It is, therefore, expected that the gap longitudinal width 
becomes greater near the rotation axis,
enhancing gap luminosity 
(per magnetic flux tube)
compared to the middle latitudes.
To see this, we must first examine the specific intensity 
of the RIAF-emitted, soft photon field in the polar funnel,
which is defined to be within $\theta<60^\circ$
in the present paper.

We tabulate the soft photon specific intensity
at each position in the magnetosphere
by the ray-tracing method.
The dispersion relation $k^\mu k_\mu=0$ gives 
the Hamiltonian,
\begin{equation}
  H
  =-k_t
  = -\frac{g_{t\varphi}}{g_{\varphi\varphi}} k_\varphi
    \pm \frac{\rho_{\rm w}}{g_{\varphi\varphi}}
        \sqrt{ k_\varphi^2
              +g_{\varphi\varphi}
               \left( g^{rr} k_r^2
                     +g^{\theta\theta}k_\theta^2 
               \right)
             } .
  \label{eq:hamilt}
\end{equation}
Thus, the Hamilton-Jacobi relation gives
\begin{equation}
  \frac{dr}{dt}
  = \frac{\partial H}{\partial k_r}
  = \pm \frac{\rho_{\rm w} n_r g^{rr}}
             {\sqrt{\lambda^2
                    +g_{\varphi\varphi}( g^{rr}n_r^2
                                      +g^{\theta\theta}n_\theta^2)}
             } ,
  \label{eq:HJ_r}
\end{equation}
\begin{equation}
  \frac{d\theta}{dt}
  = \frac{\partial H}{\partial k_\theta}
  = \pm \frac{\rho_{\rm w} n_\theta g^{\theta\theta}}
             {\sqrt{\lambda^2
                    +g_{\varphi\varphi}( g^{rr}n_r^2
                                      +g^{\theta\theta}n_\theta^2)}
             } ,
  \label{eq:HJ_th}
\end{equation}
\begin{eqnarray}
 \lefteqn{
   \frac{dn_r}{dt}
    = \frac{\partial \ln H}{\partial r}
    = \partial_r \left(\frac{g_{t\varphi}}{g_{\varphi\varphi}}
              \right)
        \lambda
         }
 \nonumber\\
 &\mp& \frac{ \displaystyle
              \partial_r\left(\frac{\rho_{\rm w}^2}
                                   {g_{\varphi\varphi}^2}
                        \right)    
                \lambda^2
             +\partial_r\left(\frac{\rho_{\rm w}^2 g^{rr}}
                                   {g_{\varphi\varphi}}
                        \right)    
                n_r^2
             +\partial_r\left(\frac{\rho_{\rm w}^2 g^{\theta\theta}}
                                   {g_{\varphi\varphi}}
                        \right)    
                n_\theta^2
            }
            {\displaystyle
             \frac{\rho_{\rm w}}{g_{\varphi\varphi}}
             \sqrt{\lambda^2
                   +g_{\varphi\varphi}( g^{rr}n_r^2
                                     +g^{\theta\theta}n_\theta^2)}
            } ,
 \nonumber\\
 \label{eq:HJ_kr}
\end{eqnarray}
\begin{eqnarray}
 \lefteqn{
   \frac{dn_\theta}{dt}
    = \frac{\partial \ln H}{\partial \theta}
    = \partial_\theta \left(\frac{g_{t\varphi}}{g_{\varphi\varphi}}
              \right)
        \lambda
         }
 \nonumber\\
 &\mp& \frac{ \displaystyle
              \partial_\theta\left(\frac{\rho_{\rm w}^2}
                                   {g_{\varphi\varphi}^2}
                        \right)    
                \lambda^2
             +\partial_\theta\left(\frac{\rho_{\rm w}^2 g^{rr}}
                                   {g_{\varphi\varphi}}
                        \right)    
                n_r^2
             +\partial_\theta\left(\frac{\rho_{\rm w}^2 g^{\theta\theta}}
                                   {g_{\varphi\varphi}}
                        \right)    
                n_\theta^2
            }
            {\displaystyle
             \frac{\rho_{\rm w}}{g_{\varphi\varphi}}
             \sqrt{\lambda^2
                   +g_{\varphi\varphi}( g^{rr}n_r^2
                                     +g^{\theta\theta}n_\theta^2)}
            } ,
 \nonumber\\
 \label{eq:HJ_kth}
\end{eqnarray}
where the wave numbers are normalized by the conserved wave energy
such that
$n_r \equiv k_r/(\hbar \omega_\infty)$ and
$n_\theta \equiv k_\theta/(\hbar \omega_\infty)$.
Note that the time coordinate $t$ for a distant static observer
plays the role of an affine parameter
because of the definition of $H$.
The initial values of $n_r$ and $n_\theta$ are calculated by
equations~(\ref{eq:LRF_redshift}),
(\ref{eq:LRF_n1}), 
(\ref{eq:LRF_n2}), 
(\ref{eq:LRF_redshift_ph}),
(\ref{eq:LRF_k1}), and
(\ref{eq:LRF_k2}).

We integrate equations~(\ref{eq:HJ_r})--(\ref{eq:HJ_kth})
along the individual rays of the RIAF-emitted photons,
and tabulate the specific intensity, $I_\omega$,
at each position on the poloidal plane in the static frame.
Note that the static limit touches the horizon at $\theta=0$,
that the inward positronic flux dominates the outward electrons'
near the horizon, and that these positrons could only tail-on collide
with the inward-unidirectional photons near the horizon. 
Thus, this treatment,
which tabulates $I_\omega$ in the static frame,
incurs only negligible errors,
although the static frame becomes space-like
near the horizon in the middle latitudes.

\subsection{The zero-angular-momentum observer (ZAMO)}
\label{sec:ZAMO}
In this paper, we compute the collision frequencies
of two-photon pair creation and inverse-Compton scatterings (ICS) 
in the frame of a zero-angular-momentum observer (ZAMO),
which rotates with the same angular frequency
as the space-time dragging frequency,
$\omega_{\rm drag} \equiv -g_{t\varphi}/g_{\varphi\varphi}$.
Putting $\beta^\varphi=\omega_{\rm drag}$
in equations~(\ref{eq:rotating_redshift}),
we thus obtain the lapse
\begin{equation}
  \frac{d\tau}{dt}
  =\alpha 
  \equiv \frac{\rho_{\rm w}}{\sqrt{g_{\varphi\varphi}}}.
  \label{eq:lapse}
\end{equation}
The tetrad of ZAMO is obtained  
from equations~(\ref{eq:rotating_e0})--(\ref{eq:rotating_e2})
and becomes
\begin{equation}
  \mbox{\boldmath$e$}_{(\hat{t})}{}^{\tiny\rm ZAMO}
  = \alpha^{-1} 
    (\mbox{\boldmath$e$}_{(t)}
     +\omega_{\rm drag}\mbox{\boldmath$e$}_{(\varphi)}),
  \label{eq:rotationg_e0z}
\end{equation}
\begin{equation}
  \mbox{\boldmath$e$}_{(\hat{r})}{}^{\tiny\rm ZAMO}
  = \sqrt{\frac{\Delta}{\Sigma}} 
    \mbox{\boldmath$e$}_{(r)},
  \label{eq:rotationg_e1z}
\end{equation}
\begin{equation}
  \mbox{\boldmath$e$}_{(\hat{\theta})}{}^{\tiny\rm ZAMO}
  = \frac{1}{\sqrt{\Sigma}} 
    \mbox{\boldmath$e$}_{(\theta)},
  \label{eq:rotationg_e2z}
\end{equation}
\begin{equation}
  \mbox{\boldmath$e$}_{(\hat{\varphi})}{}^{\tiny\rm ZAMO}
  = \frac{1}{\sqrt{g_{\varphi\varphi}}}
    \mbox{\boldmath$e$}_{(\varphi)}.
  \label{eq:rotationg_e3z}
\end{equation}

Between the distant static observer and ZAMO,
the photon energy changes by the redshift factor,
\begin{equation}
  \frac{\omega_\infty}{\omega_{\tiny\rm ZAMO}}
  =
    \frac{\mbox{\boldmath$e$}_{(t)}{}^\infty \cdot
          \mbox{\boldmath$k$}}
         {\mbox{\boldmath$e$}_{(t)}{}^{\tiny\rm ZAMO} \cdot 
          \mbox{\boldmath$k$}}
  = \frac{\alpha}{1-\beta^\varphi\lambda}.
  \label{eq:ZAMO_redshift_ph}
\end{equation}
Combining equations~(\ref{eq:LRF_redshift_ph}) and
(\ref{eq:ZAMO_redshift_ph}), we find that the 
photon energy changes from the LRF to ZAMO by the factor
\begin{equation}
  g_{\rm s} 
  \equiv \frac{\omega_{\tiny\rm ZAMO}}{\omega_{\tiny\rm LRF}}
  = \alpha^{-1} 
    \left(\frac{d\tau}{dt}\right)^{\tiny\rm LRF}
    \frac{1-\omega_{\rm drag}\lambda}{1-\Omega_{\rm K}\lambda}.
  \label{eq:redshift}
\end{equation}

\subsection{Angular distribution of RIAF soft photons in ZAMO}
\label{sec:riaf_distr}
Emitting $200 \times 128$ test photons isotropically in LRF
from $14 \times 14$ positions homogeneously 
on the poloidal plane
(\S~\ref{sec:riaf}),
we construct the specific intensity, $I_\omega$,
at each position ($r$,$\theta$)
in each directional bin in the static frame.
To compute the photon propagation direction in ZAMO,
we could use the solved $n_r$, $n_\theta$, and $n_\varphi$ and convert
the momentum from the static frame to ZAMO.
It is, however, more straightforward to use 
the photon wave vector, $k_{\hat{i}}$, measured in ZAMO.
To compute $k_{\hat{i}}$, we 
set $d\tau/dt=\alpha$ and $\beta^\varphi=\omega_{\rm drag}$ in
equations~(\ref{eq:rotating_d0})--(\ref{eq:rotating_d2})
(or equivalently, 
 eqs.~[\ref{eq:rotating_e0}]--[\ref{eq:rotating_e2}])
to obtain
\begin{eqnarray}
  k_{\hat{t}}{}^{\tiny\rm ZAMO}
    &=& -\omega_{\tiny\rm ZAMO},
  \label{eq:ZAMO_vt}\\
  k_{\hat{r}}{}^{\tiny\rm ZAMO}
    &=& b\sqrt{g_{rr}}\frac{dr}{dt},
  \label{eq:ZAMO_vr}\\
  k_{\hat{\theta}}{}^{\tiny\rm ZAMO}
    &=& b\sqrt{g_{\theta\theta}}\frac{d\theta}{dt},
  \label{eq:ZAMO_vth}\\
  k_{\hat{\varphi}}{}^{\tiny\rm ZAMO}
    &=& b\sqrt{g_{\varphi\varphi}}
                     \left( \frac{d\varphi}{dt}-\omega_{\rm drag}
                     \right),
  \label{eq:ZAMO_vph}
\end{eqnarray}
where the dispersion relation $k^\mu k_\mu=0$,
or equivalently,
\begin{equation}
  (k_{\hat r}{}^{\tiny\rm ZAMO})^2
 +(k_{\hat \theta}{}^{\tiny\rm ZAMO})^2
 +(k_{\hat \varphi}{}^{\tiny\rm ZAMO})^2
 = \omega_{\tiny\rm ZAMO}^2.
  \label{eq:ZAMO_k_i}
\end{equation}
gives
\begin{equation}
  b= \frac{\omega_{\rm ZAMO}}
          {\displaystyle
           \sqrt{ g_{rr}\left(\frac{dr}{dt}\right)^2
                 +g_{\theta\theta}
                       \left(\frac{d\theta}{dt}\right)^2
                 +g_{\varphi\varphi}
                       \left(\frac{d\varphi}{dt}
                             -\omega_{\rm drag}
                       \right)^2
                 }
          }.
  \label{eq:ZAMO_b}
\end{equation}
In the Boyer-Lindquist coordinates,
the ray-tracing result automatically gives
($dr/dt$,$d\theta/dt$,$d\varphi/dt$)
in the static frame.
Then we can readily convert it 
into the ZAMO-measured propagation direction,
($\theta_\gamma{}^{\tiny\rm ZAMO}$,
 $\varphi_\gamma{}^{\tiny\rm ZAMO}$),
using equations~(\ref{eq:ZAMO_vr})--(\ref{eq:ZAMO_vph}),
where 
\begin{eqnarray}
  k_{\hat r}{}^{\tiny\rm ZAMO}
  &=& \omega_{\tiny\rm ZAMO} \cos\theta_\gamma{}^{\tiny\rm ZAMO},
      \label{eq:ZAMO_n1}\\
  k_{\hat \theta}{}^{\tiny\rm ZAMO}
  &=& \omega_{\tiny\rm ZAMO} \sin\theta_\gamma{}^{\tiny\rm ZAMO}
      \cos\varphi_\gamma{}^{\tiny\rm ZAMO},
  \label{eq:ZAMO_n2}\\
  k_{\hat \varphi}{}^{\tiny\rm ZAMO}
  &=& \omega_{\tiny\rm ZAMO} \sin\theta_\gamma{}^{\tiny\rm ZAMO}
      \sin\varphi_\gamma{}^{\tiny\rm ZAMO}.
  \label{eq:ZAMO_n3}
\end{eqnarray}
Equation~(\ref{eq:ZAMO_n1}) gives 
the photon's propagation angle in ZAMO
with respect to the radial outward direction,
$\theta_\gamma{}^{\tiny\rm ZAMO}
 =\cos^{-1} (k_{\hat{r}}{}^{\tiny\rm ZAMO}/\omega_{\tiny\rm ZAMO})$.
Equation~(\ref{eq:ZAMO_n2}) or (\ref{eq:ZAMO_n3}) gives 
the azimuthal propagation direction, 
$\varphi_\gamma{}^{\tiny\rm ZAMO}$, 
measured around the local radial direction.

We convert the specific intensity $I_\omega$
tabulated in the static frame (\S~\ref{sec:raytrace})
into the ZAMO-measured value,
by using the invariance of $I_\omega \omega^{-3}$ 
under general coordinate transformation.
Integrating this ZAMO-measured specific intensity
over the propagation solid angle in each directional bin
at each point in ZAMO,
we obtain the soft photon flux at each point in each direction.
Finally, we use this flux to compute the ICS optical depth and
the photon-photon collision optical depth in ZAMO
\citep[see also the description around eq.~(38) of][]{li08}.

When we compute the photon-photon absorption and ICS optical depths,
we need the soft-photon differential number flux,
$dF_{\rm s}/dE_{\rm s}$
(photons per unit time per unit area per energy)
at each point and in each direction.
In \citep[][hereafter HP16]{hiro16} and H16,
we adopted the analytical solution \citep{mahad97}
of the advection-dominated-accretion flow (ADAF) as an RIAF,
and computed $dF_{\rm s}/dE_{\rm s}$
assuming that all the soft photons were emitted at $r=0$ 
with the spatially integrated ADAF spectrum
$dL/dE_{\rm s}$ 
(luminosity per photon energy),
and propagated radially to radius $r$
in a flat spacetime.
This outwardly unidirectional photon differential number flux,
$(dL/dE_{\rm s})/(4\pi r^2 E_{\rm s})$ 
(photons per unit time per unit area per energy),
is further divided by the number of 
photon-propagation-directional bins
at each point,
so that we may impose a homogeneous ADAF soft photon field.
What is more, in HP16 and H16,
we assumed that this isotropic photon differential number flux
at $r<6M$ took the same value as that at $r=6M$.
Denoting this Newtonian 
isotropic photon differential number flux at $r=6M$
as $(dF_{\rm s}/dE_{\rm s})_0$,
we can express 
$dF_{\rm s}/dE_{\rm s}=f_{\rm riaf}(dF_{\rm s}/dE_{\rm s})_0$,
where $f_{\rm riaf}=\min[1,(6M/r)^2]$.
which were used in
equations~(31), (38), and (40) of H16.

In the present paper, 
we instead compute $dF_{\rm s}/dE_{\rm s}$ in ZAMO
by the method described above.
Then the flux correction factor is tabulated at each position by
$f_{\rm riaf}(r,\theta,\cos\theta_\gamma{}^{\tiny\rm ZAMO})
 \equiv (dF_{\rm s}/dE_{\rm s})/(dF_{\rm s}/dE_{\rm s})_0$.
When computing the photon-photon absorption and ICS optical depths,
we multiply this $f_{\rm riaf}$
in the integrant of equations (38) and (40) of HP16.
Note that equation~(31) of HP16 is no longer used in the present paper,
because we abandon the mono-energetic approximation
and instead solve the Lorentz-factor dependence 
of the lepton distribution functions.

\begin{figure}
 \includegraphics[angle=0,scale=0.48]{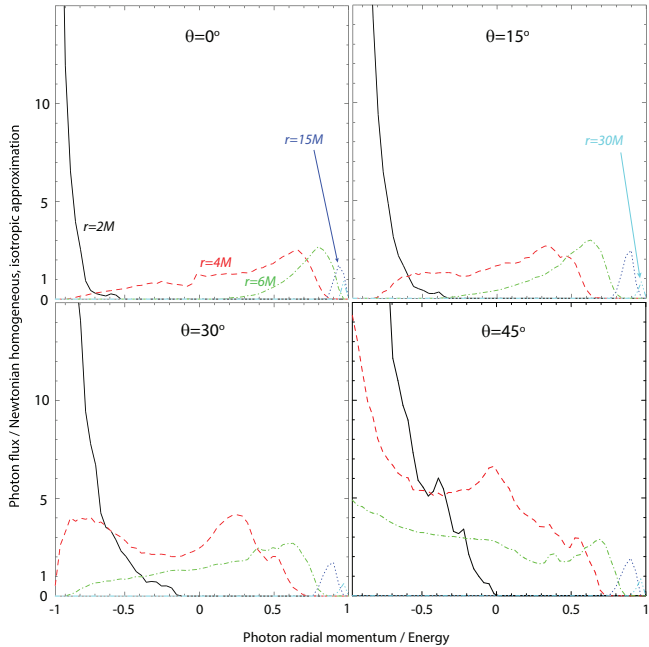}
\caption{
RIAF photon flux correction factor, $f_{\rm riaf}$,
as a function of the propagation direction,
$\theta_\gamma{}^{\tiny\rm ZAMO}$,
with respect to the radially outward direction,
at several discrete radial and latitudinal positions.
A large BH spin, $a=0.9M$, is assumed.
In the abscissa, $\cos\theta_\gamma{}^{\tiny\rm ZAMO}=-1$
corresponds to the radially inward propagation,
while $\cos\theta_\gamma{}^{\tiny\rm ZAMO}=1$
does radially outward propagation.
The black solid, red dashed, green dash-dotted,
blue dotted, and cyan dash-dot-dot-dotted curves
show $f_{\rm riaf}$
at $r=2M$, $4M$, $6M$, $15M$, and $30M$, respectively.
\label{fig:riaf}
}
\end{figure}

In figure~\ref{fig:riaf},
we present the flux correction factor,
$f_{\rm riaf}$,
assuming $a=0.9M$.
The four panels show its values along
the four discrete radial poloidal magnetic field lines,
$\theta=0^\circ$, $15^\circ$, $30^\circ$, and $45^\circ$
from the rotation axis.
The five curves in each panel represent the 
$f_{\rm riaf}$
measured at the five Boyer-Lindquist radial coordinates,
$r=2M$, $4M$, $6M$, $15M$, and $30M$
The abscissa denotes the photon propagation direction
with respect to the local radial direction:
$\cos\theta_\gamma{}^{\tiny\rm ZAMO}=-1$ corresponds to 
a radially inward propagation, while 
$\cos\theta_\gamma{}^{\tiny\rm ZAMO}=1$ 
a radially outward one.
In our previous works (HP16 and H16), 
$f_{\rm riaf}=\min[1,(6M/r)^2]$
was assumed 
in any directions $-1 < \cos\theta_\gamma{}^{\tiny\rm ZAMO} < 1$
at all the colatitude $\theta$ from the rotation axis. 

Comparing the four panels,
we find that the photon intensity decreases with 
decreasing $\theta$, because most of the photons have
positive $\lambda$ and hence difficult to approach
the rotation axis, $\theta=0^\circ$.
The solid curves (at $r=2M$) in each panel show
that the radiation field becomes
predominantly inward near the horizon, owing to the causality,
where the horizon is located at
$r_{\rm H}= 1.435 M$ in the present case of $a=0.9M$.
However, at larger $r$,
the radiation field becomes
outwardly unidirectional 
and its flux decreases by $r^{-2}$ law,
as the blue (at $r=15M$) and cyan (at $r=30M$) curves indicate.
This is because the RIAF photons are emitted only within 
$r_{\rm out}=10 M$
in the present consideration.

\section{Magnetospheric lepton accelerator near the horizon}
\label{sec:accelerator}
Being illuminated by the soft photon field described in
\S~\ref{sec:soft},
a stationary lepton accelerator can be sustained
close to the horizon (HP16).
In this section, we formulate the electrodynamics
of such a stationary BH gap,
extending the method described in H16.
In the same way as \S~\ref{sec:soft},
throughout this paper, we assume 
an aligned rotator in the sense that the magnetic axis
coincides with the rotational axis of the BH,
and consider only the \lq axisymmetric' solutions
in the sense that any quantity
does not depend on $\varphi-\Omega_{\rm F} t$.

\subsection{Magnetic field structure}
\label{sec:mag}
As demonstrated in H16, 
a stationary BH gap arises around the null surface
that is formed by the frame-dragging effect near the horizon.
Accordingly, the gap electrodynamics is essentially
governed by the frame-dragging 
rather than the magnetic field configurations.
We therefore assume a radial magnetic field on the poloidal plane,
$\Psi=\Psi(\theta)$.

Since we do not know the toroidal component of the 
magnetic field,
we cannot constrain the angular momentum of
the $\gamma$-ray photons emitted from the gap.
For simplicity, we thus assume 
that the gap-emitted photons have vanishing
angular momenta.
Under this assumption,
gap-emitted $\gamma$-rays propagate radially on the poloidal plane
and collide with the RIAF-emitted soft photons in ZAMO
with the angle 
$\theta_{\rm c} = \theta_\gamma{}^{\tiny\rm ZAMO}$
for outward $\gamma$-rays, 
and with the angle
$\pi-\theta_\gamma{}^{\tiny\rm ZAMO}$ for inward $\gamma$-rays.
To compute the rate of ICS, 
we assume that outwardly migrating {\it electrons}
collide with the soft photons with 
the same angle $\theta_{\rm c}$ as the outward $\gamma$-rays,
and that the inwardly migrating {\it positrons}
does with the same angle $\pi-\theta_{\rm c}$ 
as the inward $\gamma$-rays. 


As for the curvature process,
we parameterize the curvature radius, $R_{\rm c}$, 
of the particles, instead of constraining it from their 3-D motion.
It is, indeed, the IC-emitted, VHE photons 
(not the curvature-emitted, HE photons) that materialize
as pairs colliding with the RIAF submillimeter photons.
Thus, the actual value of $R_{\rm c}$ does not affect
the gap electrodynamics.
We thus adopt $R_{\rm c}=M$ in the present paper, 
leaving the toroidal magnetic field component, $B_\varphi$, 
unconstrained.

\subsection{Gap electrodynamics}
\label{sec:ED}
In the same way as HP16,
we solve the stationary gap solution
from the set of the Poisson equation for $\Phi$,
the equations of motion for the created leptons,
and the radiative transfer equation for the emitted photons.

\subsubsection{Poisson equation}
\label{sec:pois}
To solve the radial dependence of $\Phi$
in the Poisson equation~(\ref{eq:pois}),
we introduce the following dimensionless 
tortoise coordinate, $\eta_\ast$,
\begin{equation}
  \frac{d\eta_\ast}{dr}= \frac{r^2+a^2}{\Delta} \frac{1}{M}.
  \label{eq:tortois}
\end{equation}
In this coordinate, the horizon corresponds to 
the negative infinity, $\eta_\ast=-\infty$.
It is convenient to set $\eta_\ast=r M^{-1}$
at some large enough radius $r=r_3$.
In this paper, we put $r_3=25M$ 
(i.e., $\eta_\ast=25$ at $r=25M$),
the actual value of which never affects the results in any ways.

Since the gap is located near the horizon, 
we take the limit $\Delta \ll M^2$.
Assuming that $\Phi$ does not depend on $\varphi-\Omega_{\rm F} t$,
$\Phi=\Phi(r,\theta)$ is solved from 
the two-dimensional Poisson equation,
\begin{eqnarray}
   &&
   -\left(\frac{r^2+a^2}{\Delta}\right)^2
    \frac{\partial^2 \tilde{\Phi}}{\partial \eta_\ast{}^2}
   +\frac{2(r-M)(r^2+a^2)}{\Delta^2/M}
    \frac{\partial \tilde{\Phi}}{\partial \eta_\ast}
   \nonumber\\
   &&
   -\frac{M^2}{\Delta}
    \frac{\Sigma}{\sin\theta}
    \frac{\partial}{\partial\theta}
    \left( \frac{\sin\theta}{\Sigma}
           \frac{\partial\tilde{\Phi}}{\partial\theta} 
    \right)
    \nonumber\\
   &&
    = \left(\frac{\Sigma}{r^2+a^2}\right)^2 
      \left[ \int (n_+ -n_-) d\gamma -n_{\rm GJ}
      \right],
  \label{eq:pois_3}
\end{eqnarray}
where 
\begin{equation}
  \tilde{\Phi}(\eta_\ast) 
  \equiv \frac{1}{\Omega_{\rm F}B M^2} \Phi(r)
  \label{eq:def_Phi2}
\end{equation}
denotes the dimensionless non-corotational potential.
Dimensionless lepton distribution functions 
per magnetic flux tube are defined by
\begin{equation}
  n_\pm \equiv 
    \frac{2\pi ce}{\Omega_{\rm F} B} N_\pm(r,\theta,\gamma),
  \label{eq:def_n}
\end{equation}
where $N_+$ and $N_-$ designate the 
distribution functions of the leptons, respectively;
$\gamma$ refers to the lepton's Lorentz factor. 
Dimensionless GJ charge density per magnetic flux tube
is defined by
\begin{equation}
  n_{\rm GJ} \equiv \frac{2\pi c}{\Omega_{\rm F} B} \rhoGJ .
  \label{eq:def_nGJ}
\end{equation}
If the real charge density, $\int (n_+ -n_-) d\gamma$,
deviates from $n_{\rm GJ}$ in some region,
electric field inevitably appears along the magnetic field lines
around that region.

For a radial poloidal magnetic field, $\Psi=\Psi(\theta)$,
we can compute the acceleration electric field by
\begin{equation} 
  E_\parallel 
    \equiv -\frac{\partial \Phi}{\partial r}
    = -M \Omega_{\rm F} B
      \frac{r^2+a^2}{\Delta}
      \frac{\partial \tilde\Phi}{\partial \eta_\ast}.
  \label{eq:Ell2}
\end{equation}
Without loss of any generality, 
we can assume $F_{\theta\varphi}>0$ 
(i.e., outward magnetic field direction)
in the northern hemisphere.
In this case, a negative $E_\parallel$ arises in the gap,
which is consistent with the direction 
of the global current flow pattern. 


\subsubsection{Particle Boltzmann equations}
\label{sec:Boltzmann}
We follow the argument presented in \S~3 of \cite{hiro13}
on pulsar outer gap model.
Imposing a stationary condition, 
$\partial/\partial t
  + \Omega_{\rm F} \partial/\partial \phi = 0,
  \label{eq:stationary}
$
we obtain the following Boltzmann equations, 
\begin{equation}
  c\cos\chi  \frac{\partial n_\pm}{\partial s}
  +\dot{p}   \frac{\partial n_\pm}{\partial p}
  = \alpha(S_{{\rm IC},\pm} +S_{{\rm p},\pm}),
 \label{eq:BASIC_2}
\end{equation}
along each radial magnetic field line on the poloidal plane,
where $c$ is recovered,
the upper and lower signs correspond to the
positrons (with charge $q=+e$) and electrons ($q=-e$), respectively, and
$p \equiv \vert\mbox{\boldmath$p$}\vert=m_{\rm e}c\sqrt{\gamma^2-1}$.
Since pair annihilation and magnetic pair creation are negligible
in BH magnetospheres under low accretion rate
(and hence under weak magnetic field strength),
the right-hand side contains the collision terms
due to ICS and photon-photon pair creation.
The left-hand side is in $dt$ basis, where $t$ denotes the
proper time of a distant static observer.
Thus, the lapse $\alpha$ is multiplied in the right-hand side,
because both $S_{\rm IC}$ and $S_{\rm p}$ are evaluated in ZAMO.
On the poloidal plane, equation~(\ref{eq:metric}) gives
$ds=\sqrt{g_{rr}dr^2+g_{\theta\theta}d\theta^2}$.
However, as described in \S~\ref{sec:mag},
we neglect meridional propagation of the gap-emitted photons;
thus, we obtain $ds=dr$.

When particles emit photons via synchro-curvature process,
the energy loss ($\sim$~GeV) is small compared to the
particle energy ($\sim 10$~TeV);
thus, it is convenient to include the back reaction of 
the synchro-curvature emission
as a friction term in the left-hand side.
In this case, 
the characteristics of equation in the phase space ($s$,$p$) 
is given by 
\begin{equation}
  \dot{p} \equiv q E_\parallel \cos\chi -\frac{P_{\rm SC}}{c},
  \label{eq:char1}
\end{equation}
where the pitch angle is assumed to be
$\chi=0$ for outwardly moving positrons, and
$\chi=\pi$ for inwardly moving electrons.
In this zero-pitch-angle approximation,
the synchro-curvature radiation force 
\citep{cheng96,zhang97},
$P_{\rm SC}/c$, 
is simply given by the pure curvature formula
\citep[e.g.,][]{harding81}, 
$P_{\rm SC}/c=2e^2 \gamma^4/(3 R_{\rm c}{}^2)$.
The particle position $s$ is related with time $t$ by 
$\dot{s}= ds/dt= c \cos\chi$.

In equation~(\ref{eq:BASIC_2}), 
the IC collision terms are expressed as
\begin{eqnarray}
  S_{\rm IC} 
  &\equiv&
  -\int_{\epsilon_\gamma < \gamma}
     d \epsilon_\gamma
     \eta_{\rm IC}^\gamma(\epsilon_\gamma,\gamma,\mu_\pm) n_\pm
  \nonumber\\
  &+&
   \int_{\gamma_i > \gamma}
     d \gamma_i
     \eta_{\rm IC}^{\rm e}(\gamma_i,\gamma,\mu_\pm) n_\pm ,
  \label{eq:sec_ICa}
\end{eqnarray}
where the IC redistribution function is defined by
\begin{equation}
  \eta_{\rm IC}^\gamma 
  \equiv
    (1-\beta\mu_\pm)
    \int_{E_{\rm min}}^{E_{\rm max}}
      dE_{\rm s} 
        \frac{dF_{\rm s}}{dE_{\rm s}}
        \frac{d\epsilon_\gamma^\ast}{d\epsilon_\gamma}
        \int_{-1}^{1} d\Omega_\gamma^\ast
          \frac{d\sigma_{\rm KN}^\ast}
               {d\epsilon_\gamma^\ast d\Omega_\gamma^\ast},
  \label{eq:src_ICb}
\end{equation}    
$m_{\rm e}c^2 \epsilon_\gamma$ denotes the upscattered $\gamma$-ray
energy.
The asterisk denotes that the quantity evaluated in the
electron rest frame and 
$d\sigma_{\rm KN}^\ast/d\epsilon_\gamma^\ast d\Omega_\gamma^\ast$
denotes the Klein-Nishina differential cross section.
Energy conservation gives 
\begin{equation}
  \eta_{\rm IC}^{\rm e}(\gamma_i,\gamma,\mu_\pm)
  = \eta_{\rm IC}^\gamma(\gamma_i-\gamma,\gamma_i,\mu_\pm),  
\end{equation}
where $\gamma_i$ denotes the Lorentz factor before collision
and $\mu_+$ (or $\mu_-$) does 
the cosine of the collision angle with the
soft photon for outwardly moving electrons
(or inwardly moving positrons). 
For more details, see \S~3.2.2 of \citet{hiro03}.
The effect of inhomogeneous and anisotropic RIAF photon field
is included in the differential soft photon flux,
$dF_{\rm s}/dE_{\rm s}$,
through the correction factor $f_{\rm riaf}$
(see the last part of \S~\ref{sec:riaf_distr}).
That is, we put
$dF_{\rm s}/dE_{\rm s}=f_{\rm riaf} \cdot (dF_{\rm s}/dE_{\rm s})_0$. 
     
The photon-photon pair creation term becomes
\begin{equation}
  S_{\rm p} \equiv
    \int d\nu_\gamma
      \alpha_{\gamma\gamma}
      \frac{2\pi e}{\Omega B}
      \int \frac{I_\omega}{\hbar\omega} d\Omega,
  \label{eq:src_1b}
\end{equation}
where 
\begin{equation}
  \alpha_{\gamma\gamma}
  = (1-\mu_\pm)
    \int_{E_{\rm th}}^\infty
        dE_{\rm s} 
        \frac{dF_{\rm s}}{dE_{\rm s}}
        \frac{d\sigma_{\gamma\gamma}}{d\gamma},
  \label{eq:def_alf2}
\end{equation}
The $\gamma$-ray specific intensity
$I_\omega$ is integrated over the $\gamma$-ray propagation
solid angle $\Omega_\gamma$.
For details, see \S~3.2.2 of \citep{hiro03}.
Note that $dF_{\rm s}/dE_{\rm s}$ 
in both $\eta_{\rm IC}^\gamma$ and $\alpha_{\gamma\gamma}$
is evaluated in ZAMO as described in \S~\ref{sec:riaf_distr}.

Let us consider how the created leptons affect 
the right-hand side of equation~(\ref{eq:pois_3}).
Because $E_\parallel$ is negative,
electrons are accelerated outwards, while positrons inwards. 
As s result, 
charge density, $\int(n_+-n_-)d\gamma$, 
becomes negative (or positive)
at the outer (or inner) boundary.
In a stationary gap, $E_\parallel$ should not change sign in it.
In a vacuum gap ($n_+=n_-=0$), 
a positive (or a negative) $-n_{\rm GJ}$
near the outer (or inner) boundary makes 
$\partial_r E_\parallel >0 $ (or $\partial_r E_\parallel <0$),
thereby closing the gap.
In a non-vacuum gap, the right-hand side of equation~(\ref{eq:pois_3})
should become positive (or negative) near the outer (or inner)
boundary so that the gap may be closed.
Therefore, $\vert \int(n_+-n_-)d\gamma \vert$ should not exceed 
$\vert n_{\rm GJ} \vert$ at either boundary.
At the outer boundary, for instance, 
the dimensionless lepton distribution functions satisfy
\begin{equation}
  (n_+-n_-)\vert_{r=r_2} = -n_{-}(r_2,\gamma) 
  = j n_{\rm GJ}(r_2) \delta(\gamma-\gamma_0),
  \label{eq:BDC_n}
\end{equation}
where $\gamma_0$ denotes the Lorentz factor of the injected leptons, 
and does not affect the results unless it becomes comparable 
to the typical Lorentz factors in the gap.
The dimensionless electric current density, $j$, should be
in the range, $0 \le j \le 1$, so that the gap solution may be stationary.
If $j=1$, there is no surface charge at the outer boundary.
However, if $j<1$, the surface charge results in 
a jump of $\partial_r E_\parallel$ at the outer boundary.
That is, the parameter $j$ specifies the strength of 
$\partial_r E_\parallel$ at the outer boundary.
Thus, the inner boundary position, $r=r_1$, 
is determined as a free boundary
problem by this additional constraint, $j$.
The outer boundary position, $r=r_2$, or equivalently the gap width
$w= r_2 -r_1$, is constrained by the gap closure condition
(\S~\ref{sec:closure}).

It is noteworthy that the charge conservation ensures that
the dimensionless current density (per magnetic flux tube),
$J_{\rm e} \equiv \int (-n_+ -n_-)d\gamma$ 
conserves along the flowline.
At the outer boundary, we obtain
\begin{equation}
  J_{\rm e}
  = -\int n_-(r_2,\gamma)d\gamma
  = j n_{\rm GJ}(r_2).
  \label{eq:J_outBD}
\end{equation}
Thus, $j$ specifies not only $\partial_r E_\parallel$ at the
outer boundary, but also the conserved current density, $J_{\rm e}$.

In general, under a given electro-motive force exerted 
in the ergosphere,
$J_{\rm e}$ should be constrained by the global current 
flow pattern,
which includes an electric load at the large distances where the
force-free approximation breaks down and the trans-magnetic-field
current gives rise to the outward acceleration of charged particles 
by Lorentz forces (thereby converting the Poynting flux into
particle kinetic energies).
However, we will not go deep into the determination of $J_{\rm e}$
in this paper,
because we are concerned with the acceleration processes
near the horizon,
not the global current closure issue. 
Note that $w$ (or $r_2$) is essentially determined by $\dot{m}$;
thus, $j$ and $\dot{m}$ give the actual current density 
$(\Omega_{\rm F} B / 2\pi) J_{\rm e}$,
where $B$ should be evaluated at each position.
On these grounds, instead of determining $J_{\rm e}$ 
by a global requirement,
we treat $j$ as a free parameter in the present paper.
To consider a stationary gap solution,
we restrict the range of $j$ as $0 \le j \le 1$ 
(see the end of \S~4.2.2 of H16). 

\subsubsection{Radiative transfer equation}
\label{sec:RTE}
In the same manner as H16, 
we assume that all photons are emitted with vanishing angular momenta
and hence propagate on a constant-$\theta$ cone.
Under this assumption of radial propagation,
we obtain the radiative transfer equation \citep{hiro13}, 
\begin{equation}
  \frac{d I_\omega}{dl}= -\alpha_\omega I_\omega + j_\omega,
  \label{eq:RTE}
\end{equation}
where $dl=\sqrt{g_{rr}}dr$ 
refers to the distance interval along the ray in ZAMO,
$\alpha_\omega$ and $j_\omega$ 
the absorption and emission coefficients
evaluated in ZAMO, respectively.
We consider only photon-photon collisions for absorption,
pure curvature and IC processes for primary lepton emissions, and
synchrotron and IC processes for the emissions 
by secondary and higher-generation pairs.
For more details of the computation of 
absorption and emission,
see \S\S~4.2 and 4.3 of HP16
and \S~5.1.5 of H16. 
(Regretfully, there was a typo in an equation after eq.~[29] of H16.
 The local photon energy, $h\nu'$, is related to $h\nu$, by
 $h\nu'= \dot{t} (h\nu-m\cdot d\varphi/dt)$;
 that is, the sign should be negative, not positive.)

Some portions of the photons are emitted above 10~TeV via IC process.
A significant fraction of such hard $\gamma$-rays
are absorbed, colliding with the RIAF soft photons.
If such collisions take place within the gap, 
the created electrons and positrons polarize 
to be accelerated in opposite directions,
becoming the primary leptons.
If the collisions take place outside the gap,
the created, secondary pairs migrate along the magnetic filed lines
to emit photons via IC and synchrotron processes.
Some of such secondary IC photons are absorbed again
to materialize as tertiary pairs, 
which emit tertiary photons via synchrotron and IC processes,
eventually cascading into higher generations.

\subsubsection{Boundary conditions}
\label{sec:BDCs}
In this subsection, we describe the boundary conditions imposed 
on the three basic equations,
Eqs.~(\ref{eq:pois_3}), (\ref{eq:BASIC_2}), and (\ref{eq:RTE}),
one by one.

First, let us consider the elliptic type second-order partial differential 
equation~(\ref{eq:pois}).
In the 2-D poloidal plane,
we assume a reflection symmetry with respect to the magnetic axis.
Thus, we put $\partial_\theta \tilde\Phi=0$ at $\theta=0$.
We assume that the polar funnel is bounded at
a fixed colatitude, $\theta=\theta_{\rm max}$
and impose that this lower-latitude boundary
is equi-potential and put $\tilde\Phi=0$ at $\theta=\theta_{\rm max}$.

Both the outer and inner boundaries are treated as free boundaries.
At both boundaries, $E_\parallel$ vanishes.
To determine the positions of the two boundaries,
we impose the following two conditions:
The value of $j$ along each magnetic field line,
and the gap closure condition 
(to be described in \S~\ref{sec:closure}).
For simplicity, we assume that $j$ is constant on all the field lines.
The closure condition constrains the gap width, $w=r_2-r_1$,
and $j$ does $\partial_r E_\parallel= -\partial_r{}^2 \Phi$
at the outer boundary.
This additional condition, $\partial_r E_\parallel$
constrains the the inner boundary position $r=r_1$,
and hence the outer boundary position, $r_2=w+r_1$.

Second, consider the hyperbolic type first-order partial
differential equations~(\ref{eq:BASIC_2}).
We assume that neither electrons nor positrons 
are injected across either the outer or the inner 
boundaries.

Third, consider the first-order ordinary differential 
equation~(\ref{eq:RTE}).
We assume that photons are not injected across neither
the outer nor the inner boundaries.

\subsubsection{Gap closure condition}
\label{sec:closure}
The set of Poisson and Boltzmann equations
are solved under the boundary conditions mentioned just above.
In H16, we adopted the mono-energetic approximation
to the particle distribution functions.
However, in this paper, we explicitly solve 
their $\gamma$ dependence at each position $s$.
Accordingly, we compute the multiplicity (eq.~[41] of HP16)
of primary electrons, ${\cal M}_{\rm out}$, and that 
of primary positrons, ${\cal M}_{\rm in}$,
summing up all the created pairs by individual test particles
and dividing the result by the number of test particles.
With this modification, we apply the same closure condition 
that a stationary gap may be sustained,
${\cal M}_{\rm in} {\cal M}_{\rm out} = 1$.

\section{Gap solutions around super-massive black holes}
\label{sec:solutions}
In this paper, we apply the method described in the foregoing section
to SMBHs.
Unless explicitly mentioned,
we adopt $a=0.90 M$,
$\Omega_{\rm F}=0.50 \omega_{\rm H}$, 
$r_{\rm out}=10 M$, $j=0.7$, and
$R_{\rm c}=M$.
To solve the Poisson equation~(\ref{eq:pois_3}),
we set the meridional boundary at $\theta=\theta_{\rm max}=60^\circ$.

\subsection{The case of billion solar mass BHs}
\label{sec:billion}
Let us first examine the gap solutions
when the BH mass is $M=10^9 M_\odot$.


\subsubsection{Acceleration electric field on the poloidal plane}
\label{sec:billion_e2d}
\begin{figure}
  \includegraphics[angle=0,scale=0.70]{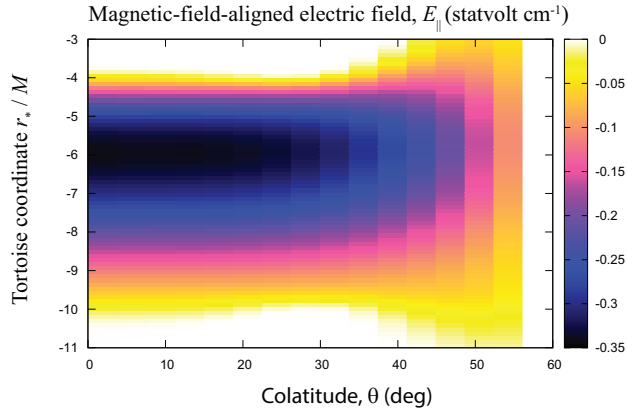}
\caption{
Acceleration electric field (statvolt ${\rm cm}^{-1}$)
on the poloidal plane.
The abscissa denotes the magnetic colatitudes, $\theta$,
in degrees,
where $0$ (i.e., the ordinate) corresponds to the magnetic axis.
The ordinate denotes the dimensionless tortoise coordinate,
where $-\infty$ corresponds to the event horizon.
The gap is solved for a black hole with $M=10^9 M_\odot$
and $a_\ast=0.9$.
Dimensionless accretion rate is chosen to be
$\dot{m}=1.77 \times 10^{-5}$.
\label{fig:e2d_1e9}
}
 \end{figure}

We begin with presenting the distribution of $E_\parallel$
on the poloidal plane.
In figure~\ref{fig:e2d_1e9},
we plot $E_\parallel$ (in statvolt ${\rm cm}^{-1}$)
as a function of
the dimensionless tortoise coordinate, 
$\eta_\ast=r_\ast/M$, and
the magnetic colatitude, $\theta$ (in degrees),
for the dimensionless accretion of $\dot{m}= 1.77 \times 10^{-5}$.
Near the lower-latitude boundary, $\theta = 60^\circ$,
a small $E_\parallel$ extends in an extended gap width 
along the poloidal magnetic field line.
However, near the rotation axis, $\theta=0^\circ$, 
much stronger $E_\parallel$ arises, 
because this region is located away from the meridional boundary
at $\theta=60^\circ$.
Since the equatorial region is assumed to be grounded 
to the near-horizon region by a dense accreting plasma, 
$E_\parallel$ vanishes in $\theta>60^\circ$. 
The selection of $60^\circ$ is, indeed, arbitrary.
For example, if the equatorial disk is geometrically thin
within $89^\circ < \theta < 91^\circ$,
figure~\ref{fig:e2d_1e9} will be stretched horizontally
to $\theta=89^\circ$.

\begin{figure}
  \includegraphics[angle=0,scale=0.40]{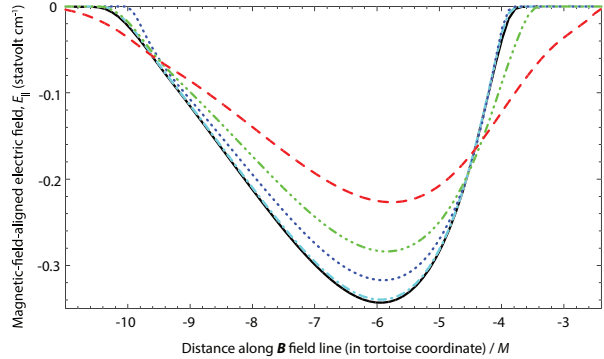}
\caption{
Acceleration electric field at five discrete colatitudes, $\theta$,
as a function of the dimensionless tortoise coordinate,
$\eta_\ast$, for $\dot{m}=1.77 \times 10^{-5}$.
The black solid, cyan dash-dotted, blue dotted, 
green dash-dot-dot-dotted, and red dashed, 
curves denote the $E_\parallel$
at $\theta= 0^\circ$, $15^\circ$, $30^\circ$, $37.5^\circ$, and
$45^\circ$, respectively. 
\label{fig:e2d_th_1e9}
}
 \end{figure}

In figure~\ref{fig:e2d_th_1e9},
we also plot $E_\parallel$ at six discrete colatitudes,
$\theta= 0^\circ$, $15^\circ$, $30^\circ$, $37.5^\circ$, and
$45^\circ$.
It follows that the $E_\parallel(\eta_\ast)$ distribution
little changes in the polar region within
$\theta \le 15^\circ$.
It is noteworthy that in H16
$E_\parallel(\eta_\ast)$ distribution little changes 
within $\theta<38^\circ$. 
The reason of this polar concentration in the present work
is that the the RIAF-emitted photons do not efficiently illuminate
the polar regions, $\theta \le 15^\circ$,
owing to their preferentially positive angular momenta.

\subsubsection{Gap emission versus colatitudes}
\label{sec:billion_SED_theta}
\begin{figure}
  \includegraphics[angle=0,scale=0.40]{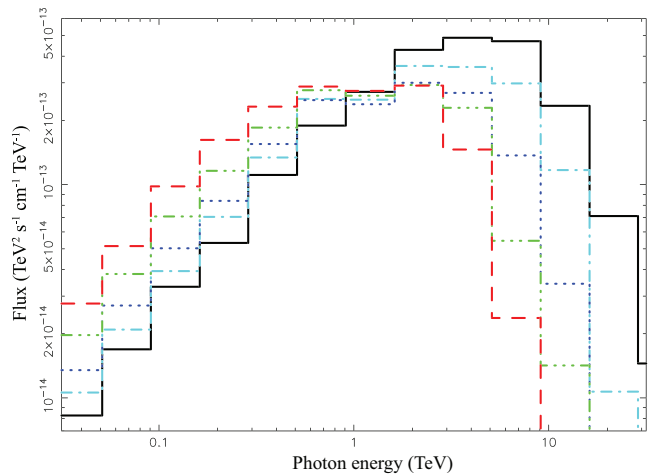}
\caption{
Spectral energy distribution (SED) of the emission from 
a black hole (BH) gap at five discrete colatitudes, $\theta$,
at a distance of 10~Mpc.
The gap is solved for a black hole with $M=10^9 M_\odot$
and $a_\ast=0.9$.
The dimensionless accretion rate is fixed at 
$\dot{m}=1.77 \times 10^{-5}$.
The five lines correspond to the same $\theta$'s as in 
figure~\ref{fig:e2d_th_1e9}.
\label{fig:SED_th_1e9}
}
 \end{figure}

We next compare the $\gamma$-ray spectra
of a BH gap emission
as a function of the colatitude, $\theta$.
In figure~\ref{fig:SED_th_1e9},
we compare the SEDs at the same five discrete $\theta$'s
as in figure~\ref{fig:e2d_th_1e9}.
It follows that the gap emission becomes
most luminous if we observe the gap
nearly face-on with a viewing angle $\theta \le 15^\circ$.
Although the $E_\parallel(s)$ distribution little changes
between $\theta=0^\circ$ and $15^\circ$,
the reduced soft photon density at $\theta \approx 0^\circ$
(particularly near the gap center, $r \approx 2M$;
 fig.~\ref{fig:riaf})
results in a smaller IC drag force and hence
greater electron Lorentz factors near the rotation axis.
Thus, the IC spectrum becomes harder at $\theta=0^\circ$
than $\theta=15^\circ$;
this point was not included in HP16 or H16,
both of which adopted a homogeneous RIAF photon density in the funnel.
It should be noted that the angular distribution of the gap emission 
is beamed into a smaller solid angle, compared to figure~4 of H16.
This is due to the small RIAF photon density near the rotation axis.

The conclusion that the gap emission gets stronger
and harder near the rotation axis, $\theta < 15^\circ$,
is unchanged if we adopt different BH masses or spins.
Therefore, in what follows, we adopt
$\theta=0$ as the representative colatitude
to estimate the greatest and hardest $\gamma$-ray flux
of BH gaps.

\subsubsection{Acceleration electric field versus accretion rate}
\label{sec:billion_Ell}
\begin{figure}
  \includegraphics[angle=0,scale=0.40]{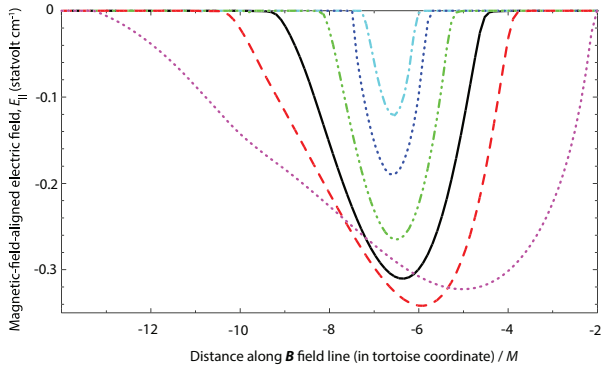}
\caption{
Spatial distribution of the magnetic-field-aligned
electric field, $E_\parallel$ (statvolt ${\rm cm}^{-1}$)
along the rotation axis, $\theta=0^\circ$,
for a black hole with $M=10^9 M_\odot$ and $a_\ast=0.9$.
The cyan dash-dotted, blue dotted, green dash-dot-dot-dotted,
red dashed, black solid, and purple dotted curves correspond to
$\dot{m}=1.78 \times 10^{-4}$,
$1.00 \times 10^{-4}$,
$5.62 \times 10^{-5}$,
$3.16 \times 10^{-5}$, 
$1.77 \times 10^{-5}$, and
$5.62 \times 10^{-6}$, respectively.
The abscissa denotes the tortoise coordinate. 
\label{fig:Ell_1e9}
}
 \end{figure}

We next consider the magnetic-field-aligned electric field,
$E_\parallel$ along the rotation axis, $\theta=0$,
as a function of the dimensionless accretion rate, $\dot{m}$.
In figure~\ref{fig:Ell_1e9},
we plot $E_\parallel(r,\theta=0)$ for six discrete $\dot{m}$'s:
the cyan, blue, green, black, red, and purple curves correspond to 
$\dot{m}=10^{-3.75}$, $10^{-4.00}$, $10^{-4.25}$, $10^{-4.50}$, 
$10^{-4.75}$, and $10^{-5.25}$.
For each case of $\dot{m}$, 
we integrate $E_\parallel$ along the poloidal magnetic field line 
to obtain the potential drop, 
$-1.87 \times 10^{15}$~V,
$-3.04 \times 10^{15}$~V,
$-6.79 \times 10^{15}$~V,
$-1.08 \times 10^{16}$~V, 
$-1.39 \times 10^{16}$~V, and
$-1.86 \times 10^{16}$~V,
respectively.
Thus, the potential drop increases (in absolute value sense)
with decreasing $\dot{m}$
because of the increased gap width, $w \equiv r_2 -r_1$.
More specifically, as the accretion rate reduces, the decreased ADAF
submillimeter photon field results in a less effective pair creation
for the gap-emitted IC photons, 
thereby increasing the mean-free path for two-photon collisions.
Since $w$ essentially becomes
the pair-creation mean-free path divided by the
number of photons emitted by a single electron
above the pair creation threshold energy \citep{hiro98},
the reduced pair creation leads to an extended gap 
along the magnetic field lines.
As a result, the smaller $\dot{m}$ is,
the greater the potential drop becomes.
It should be noted that the electro-motive force becomes
${\rm EMF}= 9.06 \times 10^{17}$~volts across the horizon
from $\theta=0^\circ$ to $60^\circ$ for $\dot{m}=5.62 \times 10^{-6}$.
That is, the potential drop in the gap attains at most 2\% of the EMF.
It is, therefore, reasonable to adopt the same $\Omega_{\rm F}$ 
inside and outside the gap along individual magnetic field lines.

As $w$ increases, the trans-field derivative 
begins to contribute in the Poisson equation~(\ref{eq:pois_3}). 
The peak of $E_\parallel$ distribution then shifts outwards,
in the same way as pulsar outer-magnetospheric gaps
\citep[fig.~12 of][]{hiro99a}.
However, the longitudinal width is at most comparable to the
perpendicular (meridional) thickness in the case of BH gaps;
thus, $E_\parallel$ does not tend to a constant value
as in pulsar outer gaps \citep{hiro06a}.

\begin{figure}
  \includegraphics[angle=0,scale=0.40]{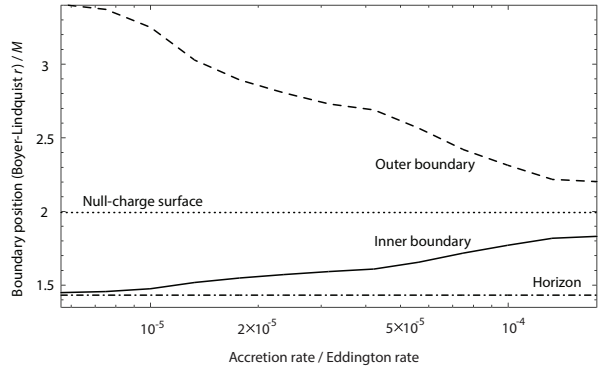}
\caption{
Spatial extent of the gap along the radial poloidal
magnetic field line
as a function of the dimensionless accretion rate.
The gap is solved for a black hole with 
$M=10^9 M_\odot$ and $a_\ast=0.9$.
Thick solid and dashed curves denote the position
of the inner and the outer boundaries of the gap,
in the unit of the gravitational radius,
$GMc^{-2}=M$. 
These boundary positions (in the ordinate) 
are transformed from the tortoise coordinate
into the Boyer-Lindquist coordinate.
The horizontal dash-dotted line shows the horizon radius,
whereas the horizontal dotted line does the null surface position
on the magnetic axis, $\theta=0$.
\label{fig:width_1e9}
}
 \end{figure}

Let us briefly examine how the gap width, $w$, is affected 
when the ADAF soft photon field changes.
In figure~\ref{fig:width_1e9}, we plot 
the gap inner and outer boundary positions 
as a function of $\dot{m}$,
where the ordinate is converted into 
the Boyer-Lindquist radial coordinate.
It follows that the gap inner boundary (solid curve, $r=r_1$), 
infinitesimally approaches the horizon
(dash-dotted horizontal line, $r=r_{\rm H}$),
while the outer boundary (dashed curve, $r=r_2$) moves outwards, 
with decreasing $\dot{m}$.
At greater accretion rate,
$\dot{m} > 1.8 \times 10^{-4}$,
we fail to find stationary solutions.
This is because only the photons that are up-scattered
in the extreme Klein-Nishina limit
can materialize as pairs in the gap,
and because the emission of such highest-energy photons 
suffers substantial
fluctuations during the Monte Calro simulation.
At smaller accretion rate, 
$\dot{m} < 5.6 \times 10^{-6}$,
there is no stationary gap solution,
because the pair creation becomes too inefficient
to create the externally imposed current density, $j$,
even when $w \gg M$.
Note that $j$ (along each magnetic flux tube)
should be constrained by a global requirement
(including the dissipative region at large distances),
and cannot be solved if we consider only the gap region.

\subsubsection{Electron distribution function}
\label{sec:billion_distr}
Because $E_\parallel$ is negative,
electrons are accelerated outwards while positrons inwards.
Thus, the outward $\gamma$-rays, which we observe,
are emitted by the electrons.
We therefore focus on the distribution function
of the electrons created inside the gap.

\begin{figure}
  \includegraphics[angle=0,scale=0.70]{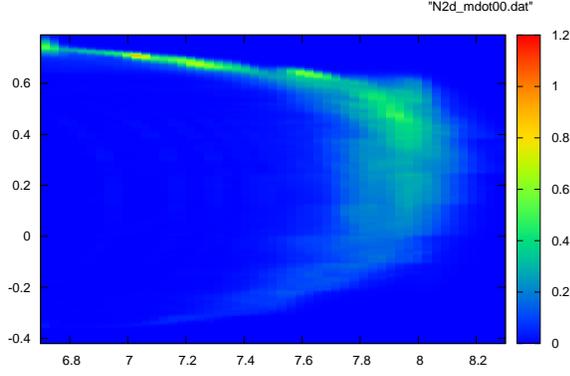}
\caption{
Distribution function, $\gamma n_-$, 
of the electrons 
along the magnetic field line on the polar axis, $\theta=0^\circ$,
for the BH with $M=10^9 M_\odot$ and $a_\ast=0.9$,
and accretion rate $\dot{m}=1.77 \times 10^{-5}$.
The abscissa denotes the electron Lorentz factor
in $\log_{10}$ scale,
while the ordinate does the distance 
from the null surface, $r-r_0$, along the poloidal 
magnetic field line.
The ordinate is converted into 
the Boyer-Lindquist radial coordinate
and plotted in gravitational radius, $M$, unit.
\label{fig:N2d_1e9}
}
 \end{figure}

In figure~\ref{fig:N2d_1e9},
we plot the electron distribution function, 
$\gamma n_-(r,\gamma)$,
along the rotation axis, $\theta=0^\circ$,
when the accretion rate is $\dot{m}=1.77 \times 10^{-5}$.
Note that $\gamma n_-$ denotes 
the electron phase-space density
per logarithmic Lorentz factor.
The abscissa denotes the Lorentz factor, $\gamma$, 
in logarithmic scale.
The ordinate denotes the distance $s$ along the magnetic field line
from the null surface, and converted into the Boyer-Lindquist 
radial coordinate (fig.~2 of H16). 
Note that $s \approx r-r_0$ holds
near the rotation axis.

Within the gap, 
electron-positron pairs are created via photon-photon collisions.
Since $E_\parallel<0$, electrons are accelerated
outward from the lower part of this figure to the upper part.
The accelerated electrons lose energy via ICS
and distribute between $4 \times 10^7 < \gamma < 1.5 \times 10^8$
in the gap.
Although a tiny $\vert\Ell\vert < 10^{-4}$,
and hence the gap extends upto $r-r_0 = 0.75M$
at $\theta=0^\circ$,
$\vert\Ell\vert$ falls down to 
$0.1 \mbox{statvolt cm}^{-1}$ at $r-r_0=0.57M$ and
$10^{-2} \mbox{statvolt cm}^{-1}$ at $r-r_0=0.68M$.
Thus, the electrons becomes nearly mono-energetic
when escaping from the gap,
forming a \lq shock' in the momentum space
(due to the concentration of their characteristics)
at $r-r_0 > 0.60 M$.

\begin{figure}
  \includegraphics[angle=0,scale=0.40]{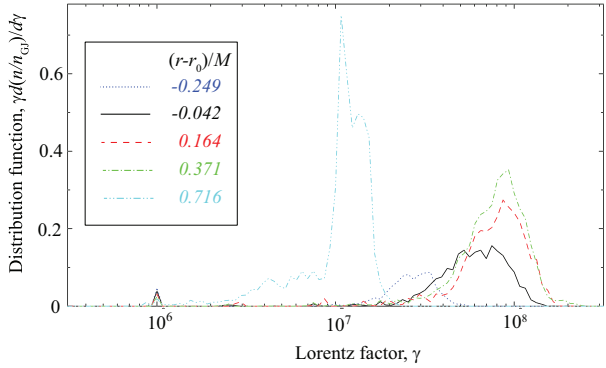}
\caption{
Distribution function, $\gamma n_-$, 
of the electrons at five positions
along the magnetic field line on the polar axis, $\theta=0^\circ$,
for the BH with $M=10^9 M_\odot$ and $a_\ast=0.9$,
and accretion rate $\dot{m}=1.77 \times 10^{-5}$.
The blue dotted, black solid, red dashed, green dash-dotted, and
cyan dash-dot-dot-dotted curves denote the $\gamma n_-$ at 
$r-r_0=-0.249M$,
$-0.042M$,
$0.164M$, 
$0.371M$, and
$0.716M$,
respectively.
\label{fig:distr_1e9}
}
 \end{figure}

In figure~\ref{fig:distr_1e9},
we also plot $\gamma n_-(r,\gamma)$ at 
five different positions,
$r-r_0=-0.249M$,
$-0.042M$,
$0.164M$, 
$0.371M$, and
$0.716M$.
Within the gap, $\gamma$ saturate below $1.5 \times 10^8$ 
because of the IC radiation drag.
Such low-energy electrons doe not efficiently emit
photons via curvature process.
Thus, the IC process dominates the curvature one.

\subsubsection{Spectrum of gap emission}
\label{sec:stellar_SED}
\begin{figure}
  \includegraphics[angle=0,scale=0.40]{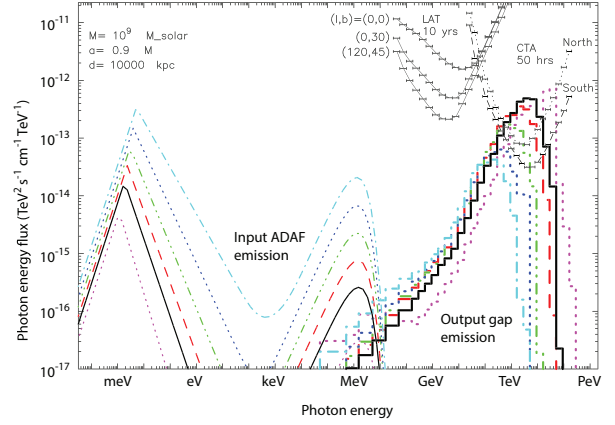}
\caption{
SED of the gap emission 
for a super-massive black hole with $M=10^9 M_\odot$ and $a_\ast=0.9$,
for six discrete dimensionless accretion rates, $\dot{m}$'s,
at 10~Mpc.
The thin curves denote the input ADAF spectra,
while the thick lines do the output gap spectra.
The cyan dash-dotted, blue dotted, green dash-dot-dot-dotted,
red dashed, black solid, and purple dotted curves correspond to
$\dot{m}=1.77 \times 10^{-4}$,
$1.00 \times 10^{-4}$,
$5.62 \times 10^{-5}$,
$3.16 \times 10^{-5}$,
$1.77 \times 10^{-5}$,
$5.62 \times 10^{-6}$, respectively.
The thin solid curves (with horizontal bars)
denote the Fermi/{\it LAT} detection limits after 10 years observation,
while 
the thin dashed and dotted curves (with horizontal bars)
denote the CTA detection limits after a 50~hours observation.
Magnetic field strength is assumed to be the equipartition value
with the plasma accretion.
\label{fig:SEDa_1e9}
}
\end{figure}

The predicted photon spectra are depicted in figure~\ref{fig:SEDa_1e9}
for six $\dot{m}$'s. 
The thin curves on the left denote the input ADAF spectra,
while the thick lines on the right do the output spectra from the gap.
We find that the emitted $\gamma$-ray flux increases 
with decreasing $\dot{m}$,
because the potential drop in the gap increases with decreasing $\dot{m}$.
The spectral peaks appear between 1~TeV and 30~TeV,
because the ICS process dominates the curvature one for 
such super-massive BHs.
The distance is assumed to be 10~Mpc.
It is clear that the gap HE flux lies well below
the detection limit of the Fermi/{\it LAT}
(three thin solid curves labeled with \lq\lq LAT 10 yrs''),
\footnote{https://www.slac.stanford.edu/exp/glast/groups/canda/\\
          lat\_Performance.htm}.
Nevertheless, its VHE flux appears above the CTA detection limits 
(dashed and dotted curves labeled with \lq\lq CTA 50 hrs'').
\footnote{https://portal.cta-observatory.org/CTA\_Observatory\\
          /performance/SieAssets/SitePages/Home.aspx}.
In the flaring state, such a large VHE flux will be
detected within one night;
thus, it is possible that a nearby 
low-luminosity, super-massive BH
exhibits a detectable gap emission above TeV
when the dimensionless accretion rate near the central BH
resides in $6 \times 10^{-6} < \dot{m} < 3 \times 10^{-5}$.

\subsection{The case of million solar masses}
\label{sec:million}
Next, let us consider a smaller BH mass
and adopt $M=10^6 M_\odot$.
To show the contribution of the curvature process
in an extended gap,
in this subsection
we consider a small accretion rate, 
$\dot{m}=1.77 \times 10^{-5}$,
which leads to a reduction of the RIAF photon field,
and hence a less effective pair creation and ICS. 
Because of the increased pair-creation mean-free path, 
the gap enlarges
from $r-r_0=-0.549M$ (i.e., almost the horizon) to
$r-r_0= 1.444M$.
What is more, because of the decreased ICS optical depth,
the curvature process becomes non-negligible
compared to the ICS one.

\begin{figure}
  \includegraphics[angle=0,scale=0.70]{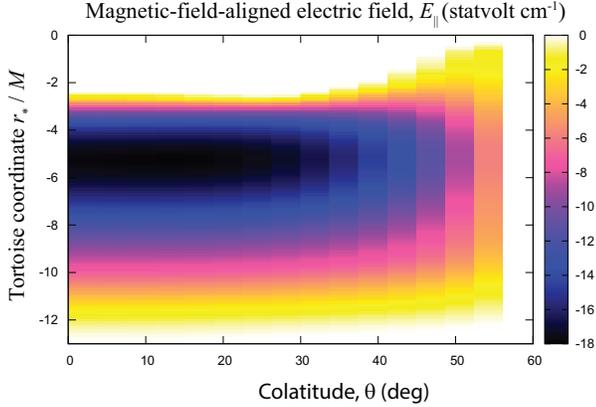}
\caption{
Similar figure as figure~\ref{fig:e2d_1e9}
but for $M=10^6 M_\odot$ 
and $\dot{m}=1.77 \times 10^{-5}$.
\label{fig:e2d_1e6}
}
 \end{figure}

In figure~\ref{fig:e2d_1e6},
we plot $E_\parallel$ (in statvolt ${\rm cm}^{-1}$)
as a function of
the dimensionless tortoise coordinate, 
$\eta_\ast=r_\ast/M$, and
the magnetic colatitude, $\theta$ (in degrees),
for the dimensionless accretion of $\dot{m}= 1.77 \times 10^{-5}$.
Near the rotation axis, $\theta=0^\circ$, 
much stronger $E_\parallel$ arises, 
in the same way as the $M=10^9 M_\odot$ case
(\S~\ref{sec:billion_e2d}).

\begin{figure}
 \epsscale{0.80}
  \includegraphics[angle=0,scale=0.40]{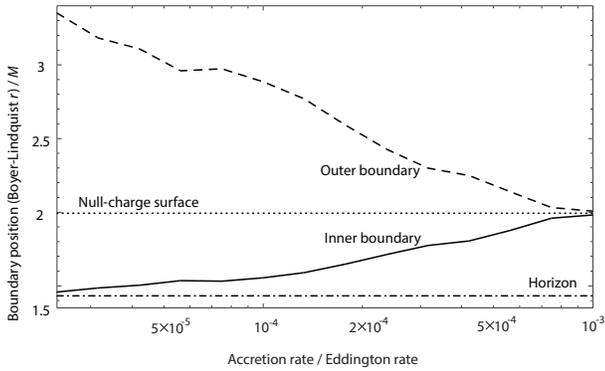}
\caption{
Similar figure as figure~\ref{fig:width_1e9}
but for $M=10^6 M_\odot$. 
\label{fig:width_1e6}
}
 \end{figure}

Let us briefly examine how the gap spatial extent depends on
$\dot{m}$, leaving from the fixed value, $1.77 \times 10^{-5}$. 
In figure~\ref{fig:width_1e6}, we plot 
the gap inner ($r=r_1$) and outer ($r=r_2$) boundary positions 
as a function of $\dot{m}$,
where the ordinate is converted into 
the Boyer-Lindquist radial coordinate.
In the same way as in the case of $M=10^9 M_\odot$,
the gap inner boundary (solid curve), 
infinitesimally approaches the horizon
(dash-dotted horizontal line),
while the outer boundary (dashed curve) moves outwards, 
with decreasing $\dot{m}$.
At smaller accretion rate, 
$\dot{m} < 2.3 \times 10^{-5}$,
there is no stationary gap solution,
because the pair creation becomes too inefficient
to create the externally imposed current density, $j$,
even when $w=r_2-r_1 \gg M$.

\begin{figure}
  \includegraphics[angle=0,scale=0.70]{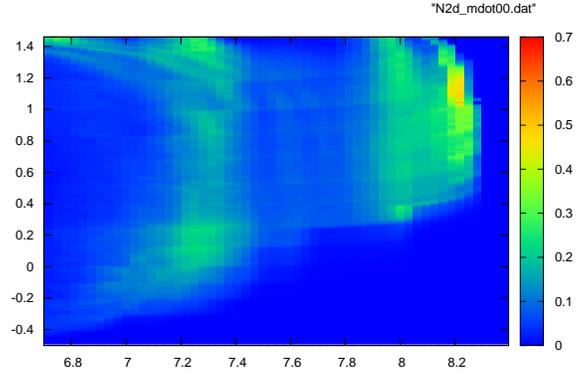}
\caption{
Similar figure as figure~\ref{fig:N2d_1e9} but for $M=10^6 M_\odot$
and $\dot{m}=1.77 \times 10^{-5}$.
The abscissa denotes the electron Lorentz factor,
while the ordinate does the distance 
from the null surface, $r-r_0$, along the poloidal 
magnetic field line in the Boyer-Lindquist radial coordinate.
\label{fig:N2d_1e6}
}
\end{figure}

Let us return to the case of $\dot{m}=1.77 \times 10^{-5}$.
In figure~\ref{fig:N2d_1e6},
we plot the electron distribution function, 
$\gamma n_-(r,\gamma)$. 
It shows that $\gamma n_-$ has a bi-modal distribution 
on $\gamma$ in the outer half of the gap, $r-r_0>0.4M$.
The lower-energy peak appears in $10^{7.2} < \gamma < 10^{7.4}$. 
Below $10^{7.2}$, ICS take place in the Thomson regime,
because the target RIAF photons are mostly infrared.
If the Lorentz factor becomes less than $10^{7.2}$, 
scattering cross section is almost unchanged, 
while energy transfer per scattering decreases 
with decreasing $\gamma$ by $\gamma^2$.
Thus, $\gamma n_-$ peaks slightly above $\gamma \sim 10^{7.2}$.
The higher-energy peak appears in 
$10^{7.85} < \gamma < 10^{8.25}$,
because the lepton Lorentz factors saturate in this range
due to the curvature radiation drag.
Unlike stellar-mass BHs (\S~5.1 of H16), however,
the curvature component contributes only mildly for SMBHs
even when $\dot{m}$ approaches its lower bound
($\dot{m}=1.77 \times 10^{-5}$ in the present case)
below which a stationary gap solution ceases to exist.

\begin{figure}
  \includegraphics[angle=0,scale=0.40]{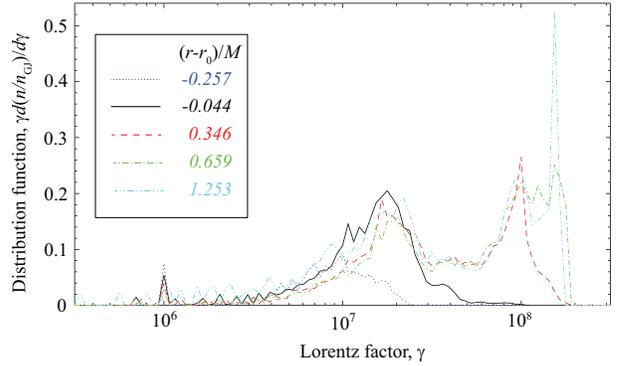}
\caption{
Similar figure as \ref{fig:distr_1e9} but for 
a black hole with $M=10^6 M_\odot$
and $\dot{m}=1.77 \times 10^{-5}$.
Other parameters are the same:
$a_\ast=0.9$ and $\theta=0^\circ$,
The blue dotted, black solid, red dashed, green dash-dotted,
and cyan dash-dot-dot-dotted curves denote the $N_-$ at 
$r-r_0=-0.257M$,
$-0.044M$,
$0.346M$, 
$0.659M$, and 
$1.253M$, 
respectively.
\label{fig:distr_1e6}
}
 \end{figure}

We plot $\gamma n(\gamma,r)$ in figure~\ref{fig:distr_1e6}
at five discrete positions, $r-r_0$, 
As electrons are accelerated, their Lorentz factors increase
as the blue dotted, black solid, and red dashed lines indicate.
When electrons escape from the gap
(cyan dash-dot-dot-dotted line),
their Lorentz factors concentrate at the terminal values
because of the concentration of the characteristics
in the momentum space.

\begin{figure}
  \includegraphics[angle=0,scale=0.40]{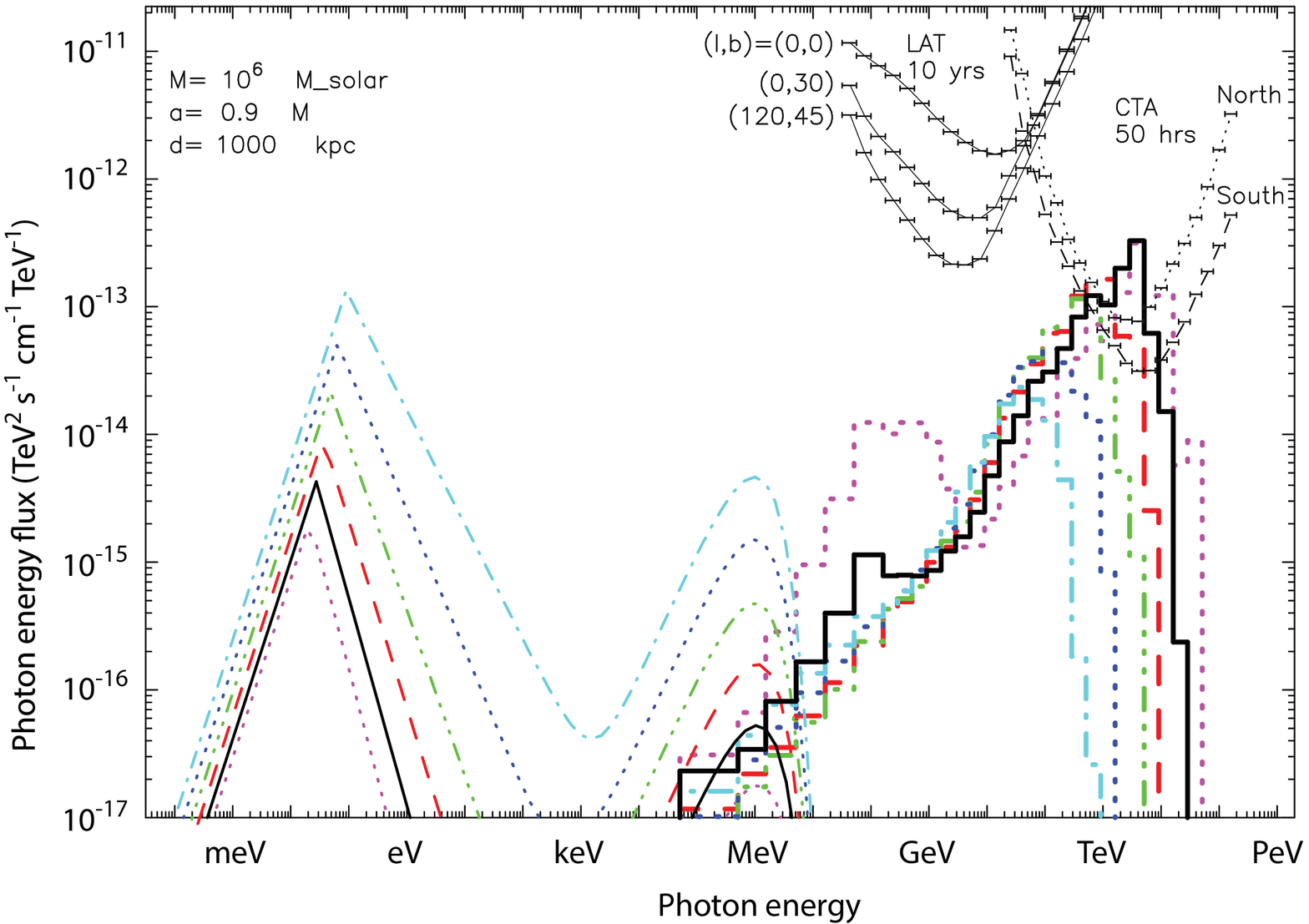}
\caption{
SED of the gap emission 
for a super-massive black hole with $M=10^6 M_\odot$ and $a_\ast=0.9$,
for six discrete dimensionless accretion rates, $\dot{m}$'s,
at 1~Mpc.
The thin curves denote the input ADAF spectra,
while the thick lines do the output gap spectra.
The cyan dash-dotted, blue dotted, green dash-dot-dot-dotted,
red dashed, black solid, and purple dotted curves correspond to
$\dot{m}=3.14 \times 10^{-4}$,
$1.00 \times 10^{-4}$,
$5.62 \times 10^{-5}$,
$4.21 \times 10^{-5}$,
$3.16 \times 10^{-5}$,
$1.77 \times 10^{-5}$, respectively.
\label{fig:SEDa_1e6}
}
\end{figure}

In figure~\ref{fig:SEDa_1e6},
we plot the predicted spectra of the gap emissions
for six discrete $\dot{m}$'s, 
assuming a luminosity distance of $1$~Mpc.
When the accretion rate is in the narrow range,
$2 \times 10^{-5} < \dot{m} < 3 \times 10^{-5}$,
we find that the gap emission will be marginally detectable 
with CTA,
particularly if the source is located in the southern sky.

\begin{figure}
  \includegraphics[angle=-90,scale=0.35]{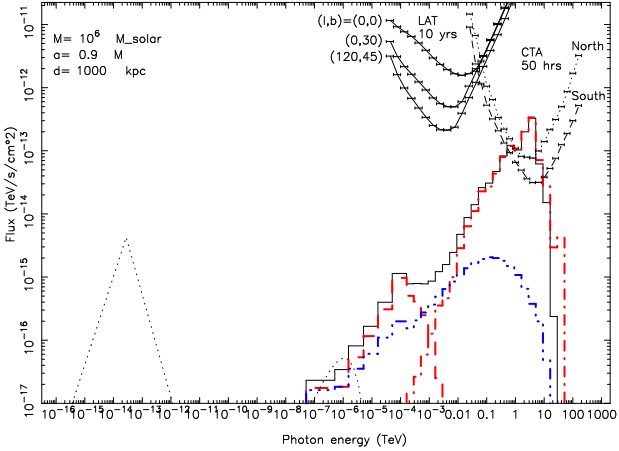}
\caption{
Similar figure as fig.~\ref{fig:SEDa_1e6},
but only the case of $\dot{m}=1.77 \times 10^{-5}$ is depicted.
The thick and thin black curves correspond to the same ones
in figure~\ref{fig:SEDa_1e6}.
The red dashed and dash-dotted lines denote
the primary curvature and inverse-Compton (IC) components, respectively.
Reprocessed, secondary IC and synchrotron components are 
are denoted by the blue dash-dot-dot-dotted curve.
\label{fig:SEDb_1e6}
}
\end{figure}

We also plot the emission components for
$\dot{m}=1.77 \times 10^{-5}$ 
in figure~\ref{fig:SEDb_1e6}.
The black solid line coincides with the purple dotted line
in figure~\ref{fig:SEDa_1e6}.
The red dash-dotted and red dashed lines represent
the IC and curvature components, respectively,
while the blue dash-dot-dot-dotted one does
the spectrum of the secondary IC and synchrotron photons
emitted outside the gap.
Because there is a population of electrons saturated at the
curvature-limited value for $\dot{m}=1.77 \times 10^{-5}$,
a weak curvature component appears between 50~MeV and a few GeV.

\section{Discussion}
\label{sec:disc}
To sum up, 
we have examined the formation of a stationary lepton accelerator 
(i.e., a gap)
in the magnetospheres of a rotating, super-massive black hole (BH).
By solving the set of an inhomogeneous part of the Maxwell equations,
lepton Boltzmann equations, and the radiative transfer equation,
we demonstrate that the null-charge surface appears
in the vicinity of a rapidly rotating BH,
and that an electric field arises along the magnetic field
line around the null charge surface,
in the same manner as in pulsar outer-gap model.
In the gap, electrons and positrons are created via two-photon collisions
and accelerated in the opposite directions 
by the magnetic-field aligned electric field 
into ultra-relativistic energies.
Such leptons emit copious $\gamma$-rays mainly via inverse-Compton (IC) 
processes, 
leading to a pair-creation cascade in the magnetosphere.
The gap longitudinal width is self-regulated so that a single electron
eventually cascades into a single pair within the gap,
and coincides, on average, with the mean-free path 
(for an IC photon to materialize via two-photon collision)
divided by the number of IC photons emitted by a single electron
above the pair-creation threshold energy.
As the accretion rate decreases,
the increased mean-free path results in an extended gap,
and hence an increased luminosity.
The gap luminosity,
which little depends on the magnetic field configuration
near the horizon, 
maximizes when the gap width becomes greater than the horizon radius.
If the BH mass is $M=10^9 M_\odot$,
these IC emissions are detectable with {\it CTA},
provided that the distance is within a few tens of Mpc
and that the dimensionless accretion rate is in the range
$6 \times 10^{-6} < \dot{m} < 3 \times 10^{-5}$.
If $M=10^6 M_\odot$,
they are detectable with {\it CTA},
provided that the distance is within a few Mpc
and that $2 \times 10^{-5} < \dot{m} < 3 \times 10^{-5}$.

\subsection{Improvement form H16}
\label{sec:disc_improvement}
In the present work, there are two major improvements from H16, 
which formulated the BH gap model and applied it to 
various BH masses in $10^1 M_\odot < M < 10^{9.8} M_\odot$.

First, 
the distribution functions of electrons and positrons
are solved as a function of the Lorentz factor, $\gamma$,
in the present work, 
whereas a mono-energetic approximation was adopted in H16.
It is found that the Lorentz factors broadly distribute below 
the saturated value that was estimated 
in the mono-energetic approximation.
This fact causes an important impact on the gap electrodynamics.
Since only the highest-energy IC photons contribute
in the gap closure (\S~\ref{sec:closure}),
the mono-energetic approximation has overestimated 
the pair creation in the gap,
thereby underestimating the gap width and luminosity
(cf. fig.~\ref{fig:SEDa_1e9} of the present paper and
 fig.23 of H16).
Moreover, the reduced Lorentz factors significantly
suppress the curvature process,
whose power is proportional to $\gamma^4$.
Also, the primary IC spectrum is softened
to peak between 1--10~TeV in the present work,
whereas it peaked between 10--100~TeV in H16.
Since CTA increases its sensitivity with decreasing photon energy
around 10~TeV,
this result encourages us to observe 
nearby low-luminosity AGNs in VHE.

Second, in the present work, 
we take into account of an anisotropic and inhomogeneous 
RIAF specific intensity in computing the IC and pair creation.
In particular, it is found that
the polar region $\theta \sim 0$ is less efficiently illuminated
by the RIAF photon field
compared to the middle-latitudes $\theta \sim 45^\circ$.
It leads to a harder and stronger VHE emission along 
the rotation axis than that into the middle latitudes.
On the contrary, in H16, 
it was simply assumed that the RIAF photon field was 
constant for $\theta$.
Thus, although the VHE flux decreased with $\theta$ 
due to the reduced $E_\parallel$ near the equatorial boundary, 
it decreased slower than the present analysis.

\subsection{Comparison with pulsar emission models}
\label{sec:disc_OG}
Let us compare the present BH gap model
with the pulsar outer-gap model.
In both gap models, the gap appears around the null-charge surface
where the GJ charge density vanishes,
as the stationary solution of the Maxwell-Boltzmann equations.
There are, however, differences as described below.

In pulsar magnetospheres,
the neutron star (NS) emits X-ray photons from its surface
losing its thermal energy. 
The luminosity of these soft photons decreases
as the NS ages.
The decreased soft photon density in the outer magnetosphere
results in an extended gap along the magnetic field line.
For middle-aged pulsars, $w$ becomes comparable to 
the radius of the outer light surface,
within which the special-relativistic GJ charge density 
changes substantially
due to the convex geometry of the magnetic field lines.
In this case, typically $20$ to $30\%$ of the NS spin-down power
is dissipated in the gap as HE $\gamma$-rays 
via the curvature process.
Note that the gap efficiency does not approach 100~\%, 
because the exerted $E_\parallel$ is less than the vacuum value
due to the partial screening by the created and separated pairs,
and because the current density is less than the GJ value.
The maximum gap power is realized when the current density
is between 50\% and 70\% of the GJ value 
(i.e., $0.5 \le j \le 0.7$).
Because the photons are emitted along the magnetic field lines
that have convex geometry,
the HE photons are emitted as a fan-like beam.

In BH magnetospheres, 
the accreting plasmas emit submillimeter photons 
from the RIAF.
Its luminosity decreases with decreasing accretion rate.
The decreases soft photon density near the horizon
(typically within a few gravitational radii for rapidly rotating BHs)
results in an extended gap width along the magnetic field line.
If $\dot{m}<10^{-4}$ for $M \sim 10^9 M_\odot$
or if $\dot{m}< 2.3 \times 10^{-5}$ for $M \sim 10^6 M_\odot$,
the gap longitudinal width becomes comparable to 
the radius of the inner light surface,
within which the GR-GJ charge density changes substantially
due to the frame-dragging.
In the same way as pulsar outer gaps, 
typically $20$ to $30\%$ of the BH spin-down power
(i.e., the BZ power)
is dissipated in the gap as VHE $\gamma$-rays 
via the IC process from such low-luminosity AGNs.
The maximum power is realized if $0.5 \le j \le 0.7$
(\S~5.1.7 of H16)
by the same reason as the pulsar outer gaps.
Because photons are preferentially emitted 
along the magnetic field lines that are nearly radial 
near the magnetic axis,
the VHE photons are emitted as a pencil-like beam,
whose geometry is similar to the pulsar polar-cap emission,
rather than the outer-gap one.

In pulsar magnetosphere,
electrons may be drawn outward as a space-charge-limited flow (SCLF)
at the NS surface in the polar-cap region.
Thus, in a stationary gap \citep{harding78,daugherty82} or 
in a non-stationary gap \citep{timo13,timo15}, 
$\gamma$-ray emission could be realized without pair creation
within the polar cap, 
although pairs are indeed created via magnetic pair creation
(e.g., at least at the outer boundary where $E_\parallel$ is screened).
However, in BH magnetospheres,
causality prevent any plasma emission across the horizon.
Thus, a gap can be sustained only with pair creation,
in the same manner as in pulsar outer gaps.
In another word, 
the BH gap electrodynamics is more close to the pulsar
outer gap rather than the polar-cap accelerator,
although its emission pattern is more close to the latter.

%
%



\acknowledgments
One of the authors (K. H.) is indebted to 
Dr. T.~Y. Saito for valuable discussion on the CTA sensitivity,
and to Dr. K. Kashiyama
for fruitful discussion. 
This work is supported by the Theoretical Institute for Advanced Research 
in Astrophysics (TIARA) operated under Academia Sinica,
and by the Ministry of Science and Technology of 
the Republic of China (Taiwan) through grants 
103-2628-M-007-003-MY3, 105-2112-M-007-033-MY2, 105-2112-M-007-002, 
103-2112-M-001-032-MY3. 
P.~H.~T.~T. receives financial support through the 
One Hundred Talents Program of the Sun Yat-Sen University.






\appendix
\section{Rotating frame of reference}
\label{sec:app_tetrad}
If a frame of reference is rotating with angular frequency
$\beta^\varphi$,
its four velocity $\mbox{\boldmath$U$}$ becomes
\begin{equation}
  \mbox{\boldmath$e$}_{(\hat{t})}
  = \mbox{\boldmath$U$}
  = \frac{dt}{d\tau}
    \left( \partial_t + \beta^\varphi \partial_\varphi 
    \right), 
  \label{eq:rotating_e0o}
\end{equation}
where $\mbox{\boldmath$e$}_{(\hat{t})}$
denotes the temporal orthonormal vector basis.
The normalization condition, 
$\mbox{\boldmath$e$}_{\hat{t}} \cdot \mbox{\boldmath$e$}_{\hat{t}}=-1$
gives the redshift factor,
\begin{equation}
  \frac{d\tau}{dt}
  =\sqrt{D},
  \label{eq:rotating_redshift}
\end{equation}
where
\begin{equation}
  D \equiv -g_{tt} -2g_{t\varphi} \beta^\varphi
           -g_{\varphi\varphi} (\beta^\varphi)^2.
  \label{eq:rotating_D}
\end{equation}
The clock in this rotating frame delays by the factor $d\tau/dt$
with respect to the distant static observer.
Putting 
\begin{equation}
  \mbox{\boldmath$e$}_{(\hat{\varphi})}
  = a_1 \partial_\varphi + b_1 \partial_t ,
  \label{eq:rotating_e3o}
\end{equation}
and imposing 
$\mbox{\boldmath$e$}_{\hat{t}} \cdot 
 \mbox{\boldmath$e$}_{\hat{\varphi}}=0$
and 
$\mbox{\boldmath$e$}_{\hat{\varphi}} \cdot 
 \mbox{\boldmath$e$}_{\hat{\varphi}}=1$,
we obtain
\begin{equation}
  a_1= -\frac{g_{tt}+g_{t\varphi}\beta^\varphi}
             {\rho_{\rm w} \sqrt{D}}
  \label{eq:rotating_a1}
\end{equation}
and 
\begin{equation}
  b_1=  \frac{g_{t\varphi}+g_{\varphi\varphi}\beta^\varphi}
             {\rho_{\rm w} \sqrt{D}}.
  \label{eq:rotating_b1}
\end{equation}
For completeness, we also write the radial 
and meridional orthonormal bases:
\begin{equation}
  \mbox{\boldmath$e$}_{(\hat{r})}
  =\sqrt{g^{rr}} \partial_r,
  \label{eq:rotating_e1o}
\end{equation}
\begin{equation}
  \mbox{\boldmath$e$}_{(\hat{\theta})}
  =\sqrt{g^{\theta\theta}} \partial_\theta,
  \label{eq:rotating_e2o}
\end{equation}

We can inversely solve equations~(\ref{eq:rotating_e0o})
--(\ref{eq:rotating_e2o}) to obtain the following 
coordinate bases:
\begin{equation}
  \mbox{\boldmath$e$}_{(t)}
  \equiv \partial_t
  =-\frac{d\tau}{dt}
    \frac{g_{tt}+g_{t\varphi} \beta^\varphi}
         {D}
    \mbox{\boldmath$e$}_{(\hat{t})}
   -\frac{\rho_{\rm w}}{\sqrt{D}}
    \beta^\varphi
    \mbox{\boldmath$e$}_{(\hat{\varphi})},
  \label{eq:rotating_e0}
\end{equation}
\begin{equation}
  \mbox{\boldmath$e$}_{(\varphi)}
  \equiv \partial_\varphi
  =-\frac{d\tau}{dt}
    \frac{g_{t\varphi}+g_{\varphi\varphi} \beta^\varphi}
         {D}
    \mbox{\boldmath$e$}_{(\hat{t})}
   +\frac{\rho_{\rm w}}{\sqrt{D}}
    \mbox{\boldmath$e$}_{(\hat{\varphi})},
  \label{eq:rotating_e3}
\end{equation}
\begin{equation}
  \mbox{\boldmath$e$}_{(r)}
  \equiv \partial_r
  =\sqrt{g_{rr}} \mbox{\boldmath$e$}_{(\hat{r})},
  \label{eq:rotating_e1}
\end{equation}
\begin{equation}
  \mbox{\boldmath$e$}_{(\theta)}
  \equiv \partial_\theta
  =\sqrt{g_{\theta\theta}} \mbox{\boldmath$e$}_{(\hat{\theta})}.
  \label{eq:rotating_e2}
\end{equation}

If necessary, we may use the dual one-form bases. 
From equations~(\ref{eq:rotating_e0o}),
(\ref{eq:rotating_e3o}), (\ref{eq:rotating_e1o}) and
(\ref{eq:rotating_e2o}), we obtain
\begin{equation}
  \tilde{\omega}^{(\hat{t})}
  = -\frac{d\tau}{dt}
     \left( \frac{g_{tt}+g_{t\varphi} \beta^\varphi}
                 {D}
            \tilde{d}t
           -\frac{g_{t\varphi}+g_{\varphi\varphi} \beta^\varphi}
                 {D}
            \tilde{d}\varphi
     \right),
  \label{eq:rotating_d0}
\end{equation}
\begin{equation}
  \tilde{\omega}^{(\hat{\varphi})}
  = \frac{\rho_{\rm w}}{\sqrt{D}}
    \left( \tilde{d}\varphi
          -\beta^\varphi \tilde{d}t
    \right),
  \label{eq:rotating_d3}
\end{equation}
\begin{equation}
  \tilde{\omega}^{(\hat{r})}
  = \sqrt{g_{rr}} \tilde{d}r ,
  \label{eq:rotating_d1}
\end{equation}
\begin{equation}
  \tilde{\omega}^{(\hat{\theta})}
  = \sqrt{g_{\theta\theta}} \tilde{d}\theta ,
  \label{eq:rotating_d2}
\end{equation}

\end{document}